\documentclass{aa} 

\def\void#1{{}}

\def\j{$J$}
\def\h{$H$}
\def\k{$K_s$}

\usepackage{graphicx}
\usepackage{times}
\usepackage{mathptm}
\usepackage{amssymb}
\usepackage{amsmath}
\fontfamily{ptmcm}\selectfont
\usepackage{natbib}

\begin{document}

\title{ESO Imaging Survey: Infrared observations of CDF-S
  and HDF-S\thanks{Based on
observations carried out at the European Southern Observatory, La
Silla, Chile under program Nos. 61.E-9005, 162.O-0917, 163.O-0740}}
\author{
 L. F. Olsen\inst{1,2,3}
 \and J.-M. Miralles\inst{2,4}
 \and L. da Costa\inst{2,5}
 \and C. Benoist\inst{1,2}
 \and B. Vandame\inst{2}
 \and R. Rengelink \inst{2}
 \and C. Rit\'e\inst{2,5}
 \and M. Scodeggio \inst{6}
 \and R. Slijkhuis\inst{2}
 \and A. Wicenec\inst{2}
 \and S. Zaggia \inst{7}
 }

\institute{
Observatoire de la C\^ote d'Azur, Laboratoire Cassiop\'ee, BP4229, 06304 Nice Cedex 4, France
\and
European Southern Observatory, Karl-Schwarzschild-Str. 2, D-85748
Garching b. M\"unchen, Germany
\and Dark Cosmology Centre, Niels Bohr Institute, University of Copenhagen, Juliane Maries Vej 30, DK-2100 Copenhagen,
      Denmark
\and T\`ecniques d'Avantguarda, Avda. Carlemany 75, AD-700 Les Escaldes, Andorra
\and Observat\'orio Nacional, Rua Gen. Jos\'e Cristino 77, Rio de Janeiro, R.J., Brasil
\and Istituto di Fisica Cosmica - CNR, Milano, Italy
\and Osservatorio Astronomico di Trieste, Via G.B. Tiepolo 11, I-31144
      Trieste, Italy
}
\offprints{L.F. Olsen, LisbethFogh.Olsen@obs-nice.fr}

\date{Received -; Accepted -}


   \abstract {This paper presents infrared data obtained from
observations carried out at the ESO 3.5m New Technology Telescope
(NTT) of the Hubble Deep Field South (HDF-S) and the Chandra Deep
Field South (CDF-S).  These data were taken as part of the ESO Imaging
Survey (EIS) program, a public survey conducted by ESO to promote
follow-up observations with the VLT. In the HDF-S field the infrared
observations cover an area of $\sim$ 53 square arcmin, encompassing
the HST WFPC2 and STIS fields, in the $JHK_s$ passbands. The seeing
measured in the final stacked images ranges from 0\farcs79 to
1\farcs22 and the median limiting magnitudes (AB system, 2'' aperture,
5 $\sigma$ detection limit) are $J_{AB}\sim23.0$, $H_{AB}\sim22.8$ and
$K_{AB}\sim23.0$~mag. Less complete data are also available in $JK_s$
for the adjacent HST NICMOS field.  For CDF-S, the infrared
observations cover a total area of $\sim$ 100 square arcmin, reaching
median limiting magnitudes (as defined above) of $J_{AB}\sim23.6$ and
$K_{AB}\sim22.7$~mag. For one CDF-S field \h ~band data are also
available.  This paper describes the observations and presents the
results of new reductions carried out entirely through the
un-supervised, high-throughput EIS Data Reduction System and its
associated EIS/MVM C++-based image processing library developed, over
the past 5 years, by the EIS project and now publicly available. The
paper also presents source catalogs extracted from the final co-added
images which are used to evaluate the scientific quality of the survey
products, and hence the performance of the software.  This is done
comparing the results obtained in the present work with those obtained
by other authors from independent data and/or reductions carried out
with different software packages and techniques. The final
science-grade catalogs together with the astrometrically and
photometrically calibrated co-added images are available at
CDS\thanks{The science grade images and catalogs are available at CDS
via anonymous ftp to cdsarc.u-strasbg.fr (130.79.128.5) or via
http://cdsweb.u-strasbg.fr/cgi-bin/qcat?J/A+A/}.

\keywords{catalogs -- surveys -- stars:general -- galaxies:general --
                cosmology:observations}} \maketitle

%

\section{Introduction}

One of the goals of the ESO Imaging Survey \citep[EIS, ][]{renzini97}
was to carry out moderately deep, multi-passband, optical/infrared
observations over relatively large areas to produce faint galaxy
samples. The primary objective was to produce samples with color
information to estimate photometric redshifts, extract statistical
samples of galaxies likely to be in the poorly sampled $1
\lesssim z<2$ redshift interval or  Lyman-break candidates at
$z\gtrsim2.5$ and, perhaps, examples of high redshift quasars and
galaxies all of which are interesting targets for follow-up
spectroscopic observations with the VLT.

The first such attempt was the EIS-DEEP survey carried out at the
NTT in the period 1998-2000. The infrared part of the survey was
carried out in the period 1998-1999 and is reported in the present
paper.  While by recent standards this survey is relatively modest in
size, the combination of depth and area made it unique at the time it
was conceived. The survey consisted of optical and infrared
observations of the Hubble Deep Field South
\citep[HDF-S, ][]{williams00} and the Chandra Deep Field South
\citep[CDF-S, ][]{giacconi01}. The choice of the regions was an attempt
by the EIS Working Group to re\-concile the great interest generated by
the HST observations of a  region selected in the southern hemisphere
(HDF-S), with the desire to conduct observations in a less crowded
region and, in particular, devoid of bright stars. Therefore, the
second region was chosen to overlap the area, selected for its low HI
column density, for which deep Chandra observations were already being
planned. This region is particularly well-suited for ground-based
imaging due to its lack of bright stars. Not surprisingly, it
eventually became a widely explored window to the high redshift
Universe with deep multi-wavelength observations being conducted by
all major space observatories and several ground-based follow-up
programs \citep[e.g. ][]{giavalisco04}.

Preliminary results of this survey were originally reported in
preprints by \cite{dacosta98} and \cite{rengelink98} who presented
fully calibrated optical and infrared images, single-passband catalogs
extracted from these images, color catalogs based on a reference
$\chi^2$ image and lists of color-selected high-redshift galaxy
candidates utilizing the Ly-break technique. All of these data
products were made publicly available world-wide. The optical images
were reduced using IRAF and the infrared images using the program {\it
Jitter} of the {\it Eclipse} library \citep{devillard99} developed by
ESO to reduce ISAAC and SOFI data.

Despite satisfactory attempts to assess the quality of the derived
products by, for instance, comparing counts of galaxies with those of
other authors and of star counts and the stellar locus in color-color
diagrams with model predictions, subsequent comparisons with deeper
data and with independent analysis using other reduction techniques
showed that the reductions carried out by the EIS team using the
available {\it Jitter} package led to a systematic loss of flux at
faint magnitudes. The color catalogs released also showed some
unexpected results which, after some time, were traced to relative
offsets between the astrometric calibration of the optical and
infrared images. Due to these problems the abovementioned
preprints were withdrawn and the data release described in the present
work supersedes those made earlier by the EIS team of this dataset.
The identified problems, together with the need to develop a common,
high-throughput system to handle optical/infrared, single and
multi-chip instruments, led to the development, as part of the EIS
project, of an image processing engine \citep[EIS/MVM library,
][]{vandame04}. This library is a central part of the integrated,
end-to-end EIS Data Reduction System \citep{dacosta04} built
to carry out un-supervised image reductions, stacking and catalog
preparation.  Preliminary results obtained with the new image
processing library, in particular addressing the flux loss problem
were originally reported by \cite{vandame01}. However, even though a
significant improvement was achieved, additional testing indicated
that an error was introduced in the code. To avoid additional
intermediate releases, the final release of these data was postponed
until the completion of the EIS Data Reduction System infrastructure
which includes the photometric pipeline and provides the
infrastructure required for re-processing and re-calibrating data in
an efficient way, a key element for testing new software.

In this paper these new reductions of the data accumulated by
EIS-DEEP, including new ones not previously released, are
presented. In Sect.~\ref{obs}, the observations, including those
carried out in 1999 and not previously reported, are reviewed. 
The goal of this paper is two-fold: first, provide the most recent
reductions and substitute previous releases; second, use these data as
a test case to justify the use of the developed software to treat the
much larger dataset accumulated by the infrared part of the Deep
Public Survey (DPS) carried out by the EIS team over 81 nights in the
period 2000-2004, which generated about 15000 science frames. The
results for this survey will be presented in a forthcoming paper of
this series \citep{olsen06b}. In Sect.~\ref{reduction} the procedures
adopted to remove the instrumental signature and to astrometrically
and photometrically calibrate the nightly data are discussed, while in
Sect.~\ref{sec:finalproducts} the attributes of the final co-added
images and their associated catalogs are presented.  The scientific
quality of these advanced survey products is assessed by comparing in
Sect.~\ref{discussion} the results derived from the present dataset
with those obtained from a variety of other datasets reduced using
diverse techniques. The main results of the present paper are
summarized in Sect.~\ref{summary}.

\section {Observations}
\label{obs}

\begin{table}[t]
\center
\caption{SOFI Pointings (J2000.0).}
\label{tab:pointings}
\begin {tabular}{lcccc}
\hline \hline
Field &  $\alpha$ & $\delta$ \\
\hline
HDF-S-1 & 22:33:29.1 & -60:33:50 \\
HDF-S-2 & 22:32:42.4 & -60:33:50 \\
HDF-S-3 & 22:33:00.0 & -60:37:59 \\
\hline
CDF-S-1 &  03:32:16.7 & -27:46:00 \\
CDF-S-2 &  03:32:38.0 & -27:46:00 \\
CDF-S-3 &  03:32:16.7 & -27:50:35 \\
CDF-S-4 &  03:32:38.0 & -27:50:35 \\
\hline
\end{tabular}
\end{table}

\begin{table}
\caption{Number of observed fields in each region.}
\label{tab:obs}
\center
\begin{tabular}{lcccc}
\hline\hline
Region & $J$ & $H$ & $K_s$ \\
\hline
CDF-S &  4 & 1 & 4\\
HDF-S &  3 & 2 & 3\\
\hline
\end{tabular}
\end{table}

As described above, the EIS-DEEP observations originally were
planned to cover two regions: one region consisting of three adjacent
fields, of about 25 square~arcmin each, covering the WFPC2, STIS and
NIC3 fields; and another region constituting a $2\times2$ mosaic,
comprising a total area of $\sim100$~square arcmin covering part of
the now well-known CDF-S region.  The center of the mosaic was chosen
to coincide with the nominal position selected for the X-ray
observations, where the sensitivity and image quality were expected to
be the best.  The pointings adopted in the construction of the mosaics
are listed in Table~\ref{tab:pointings}.

The observations were obtained using the SOFI camera \citep{moorwood98}
mounted on the NTT. SOFI is equipped with a Rockwell 1024$^2$ detector
that, when used together with its large field objective, provides
images with a pixel scale of 0.29~arcsec, and a field of view of about
$4.9\times4.9$~square arcmin. The total area covered  in $J$ and
$K_s$ is about $\sim175$~square arcmin split into two mosaics,
one of $\sim75$~square arcmin covering HDF-S and the other of
$\sim100$~square arc\-min covering CDF-S. The observations were
carried out in groups of exposures (OBs). The excution of an OB
consists of a series of short exposures with small position offsets
from the target position.  The observations were conducted using the
{\tt AutoJitter} observing template, which implements the offsetting
of adjacent exposures within a box of a user specified size $s$,
chosen to be $s=45$~arcsec, approximately 15\% of the field of view.
The offsets are constrained so that all distances between pointings,
in a series of 15 consecutive exposures are larger than 9~arcsec. Each
observation comprised sixty one-minute exposures, which themselves
were the average of six ten-second sub-exposures. Table~\ref{tab:obs}
summarizes the available observations for each of the mosaics. The
numbers listed in the table represent the number of observed
fields regardless of their exposure times.

A total of 3051 science frames were obtained. Out of
those, 1316 are in the CDF-S region and represent 22 hours of
on-target integration, obtained during 7 nights in the period
August 3, 1998 and November 11, 1998. The remaining  1735 are
located in the HDF-S region and were accumulated in 29 hours of
on-target integration, in observations carried out during 14
nights over the period August 3, 1998 to July 7, 1999. Observations
of standard stars are available for all nights, generating 613
images taken in 7805 sec.

\section {Data reduction}
\label{reduction}

The new reduction of the data, described in the present paper, was
carried out using the EIS Data Reduction System \citep{dacosta04},
which is an end-to-end, integrated system providing the infrastructure
for the un-supervised reduction and calibration of large volumes of
imaging data. An integral part of the reduction system is a C++-based
image processing library \citep[EIS/MVM library, ][]{vandame04}
developed as part of the EIS effort. It is both a library of basic
image processing routines, and an integrated, high-throughput pipeline
(reaching a sustainable rate of about 0.5~Mpix/sec) for automated
reduction of single or multi-chip optical/infrared images.

The main steps of the EIS Data Reduction System are: the creation of
reduced images on a night by night basis; the photometric calibration
of the reduced images; stacking of the reduced images yielding the
final stacked images forming the basis of the catalog extraction.  The
EIS/MVM pipeline is responsible for the first step of producing fully
reduced images and weight-maps carrying out bias subtraction,
flatfielding, de-fringing, correction of eletronic cross-talk effects,
identification and automatic masking of satellite tracks, and
sky-background subtraction as well as first-order pixel-based image
stacking (allowing for translation, rotation and stretching of the
image). The stacking process combines the reduced images into final
image stacks properly flux scaling the images to conserve the
photometric calibration. Finally, the object catalogs are extracted
using SExtractor.  The reductions of optical wide-field imaging data
are presented by \citet[ hereafter Paper~I]{dietrich05} and
\cite{mignano06}, where details about common aspects of the reduction
can be found.  Paper~I also includes a detailed description of the
released images and catalogs and their format. Here only the more
important steps for handling infrared data are reviewed.

After the basic image reduction, the de-fringing, sky-subtraction and
elimination of image artifacts are carried out in a two-step
procedure: first, a preliminary image is obtained by stacking a set of
sky-subtracted frames, with the background estimated using a running
median; second, objects are identified in this stacked image and masks
are created and enlarged by a user-specified factor; finally, the
de-fringing/sky-subtraction step is repeated with the object masks
mapped to the jittered position of each individual image. In the final
co-addition of the images the masked pixels are neglected and deviant
pixels are clipped. This procedure is particularly important for
infrared observations since, as explained in Sect.~\ref{obs}, the
observations are carried out as sequences of short exposures, in order
to allow a suitable estimate of the fast varying background at these
wavelengths to be made.  The short exposure time of the individual
images, prevent the detection of a large number of objects with
non-negligible flux, which can make a significant contribution to the
background when several images are co-added.  Therefore, if these
objects are not properly masked when estimating the background, the
background is over-estimated, biasing the flux of even relatively
bright objects. This was the effect observed in the original release
of the EIS-DEEP data using the {\it Eclipse} reduction package
\citep{dacosta98, rengelink98}. With the two-step procedure now
adopted, the flux estimates have greatly improved and, as shown in
Sect.~\ref{discussion}, the flux estimates are in excellent agreement
with those obtained using other reduction packages commonly used
to reduce infrared data.

The library also includes advanced algorithms used for the astrometric
calibration and warping of reduced images onto a user-defined
reference grid.  The astrometric calibration of the infrared frames
was performed using the GSC2.2 (HDF-S) and USNO-B \citep[CDF-S,
][]{monet03} catalogs as reference following the method developed by
\cite{djamdji93}.  The GSC2 catalog was used because of the lack of
reference stars in the USNO-B within the surveyed are of the HDF-S
region. The internal astrometric accuracy is about half a pixel
($\sim$0.15~arcsec) and is limited by the internal accuracy of the
reference catalog. When dealing with multi-wavelength surveys it is
preferable to register the infrared data using a catalog from an
optical counterpart image.  This will not only provide a much larger
number of reference objects but also avoid the effects associated with
small deviations of the astrometric solutions obtained for the optical
and infrared data that were identified in the original release of the
EIS-DEEP data.

The photometric calibration for each night is based on observations of
standard stars taken from \cite{persson98}.  Typically two to seven
stars were observed over a range of airmass. For all nights,
independent photometric solutions were attempted using the photometric
pipeline of the EIS Data Reduction System. Depending on the airmass
and color coverage linear fits with  one, two or three free
parameters are attempted. If no standard star observations are
available or no suitable fits are possible a default solution is
adopted.  This was the case for only two nights: August 4, 1998 for
both \j\ and \k\ bands and November 11, 1998 for the \k-band with
insufficient observations of standard  stars during the night to
derive a reliable solution.

\begin{table}
\caption{Type of photometric solutions available for each combination of mosaic and band.}
\label{tab:ppsolutions}
\center
\begin{tabular}{lcccccc}
\hline
Region & Passband & Default & 1-par & 2-par & 3-par & total \\
\hline\hline
CDF-S  &   $J$   & 0 &  1 & 2 & 0 & 3  \\
CDF-S  &   $H$   & 0 &  0 & 1 & 0 & 1 \\
CDF-S  &   $K_S$ & 1 &  1 & 2 & 0 & 4 \\
HDF-S  &   $J$   & 1 &  4 & 0 & 0 & 5 \\
HDF-S  &   $H$   & 0 &  4 & 0 & 0 & 4 \\
HDF-S  &   $K_s$ & 1 &  4 & 1 & 0 & 6 \\
\hline
\end{tabular}
\end{table}

\begin{table}
\caption{Average photometric solutions. There was no case where
the color term could be determined.}
\label{tab:photsol}
\center
\begin{tabular}{ccc}
\hline\hline
Passband & Zp & k\\
\hline
\j & $23.21\pm0.05$ & $0.11$\\
\h & $23.00\pm0.05$ & $0.02$\\
\k & $22.45\pm0.06$ & $0.13$\\
\hline
\end{tabular}
\end{table}

\begin{table}
\caption{Comparison between EIS and Telescope Team solutions
(EIS-Telescope). There was no case where the color term could be
determined.}
\label{tab:photsol_comp}
\center
\begin{tabular}{ccr}
\hline\hline
Passband & Zp & k\\
\hline
\j & -0.07 & 0.01\\
\h & -0.06 & -0.02\\
\k & 0.0 & 0.03\\
\hline
\end{tabular}
\end{table}

Table~\ref{tab:ppsolutions} summarizes the type of photometric
solutions available for each mosaic. The table gives in Col.~1 the
region name; in Col.~2 the passband; in Col.  3 the number of
default  solutions;
in Cols. 4--6 the number of nights with photometric solutions derived
from 1- to 3- parameter fits; and in Col.~7 the total number of nights
with science and standard star observations for a given combination of
mosaic and passband. It can be seen that in many cases only
1-parameter fits are available, in particular in the HDF-S mosaic, and
that no solution was obtained with an independent estimate of the
color term. This is the result of insufficient airmass and/or color
coverage of the standard star observations. Note that for mosaics, in
principle, one good solution suffices to determine a reliable
calibration, using overlapping regions to propagate a given solution.

Table~\ref{tab:photsol} summarizes the average photometric solutions
determined for the entire period of observations, covering more than
one year. The table lists in Col.~1 the passband; in Col.~2 the mean
zeropoint and standard deviation; in Col.~3 the mean extinction.  The
number of solutions is in general small ranging from 1 to 6 solutions
in the different bands and only for 1--3 of these could the extinction
be determined, as can also be seen from
Table~\ref{tab:ppsolutions}. Therefore, the standard deviation for the
extinction values could not be determined reliably. The errors
estimated for the extinction values are typically of the order of
$\lesssim0.05$. The listed standard deviations are a good indication
of the uncertainty of the estimated zeropoints. The solutions have
been compared with the values provided by the Telescope
Team\footnote{http://www.ls.eso.org/lasilla/sciops/ntt/sofi/setup/Zero\_Point.html}
in Table~\ref{tab:photsol_comp} listing in Col.~1 the passband; in
Col.~2 the difference in zeropoint of the photometric solutions
computed as (EIS-Telescope Team); in Col.~3 the corresponding
difference for the determined extinction. It can be seen that the
differences are all within the estimated uncertainty, which is
noteworthy given that the solutions are not derived for the same
period of time.  Below, the derived magnitudes are compared with those
of other authors, and thus careful consideration is given to the
photometric quality of the data. Where the magnitudes are listed in
the $AB$ system the following conversions have been used: $J_{AB}= J +
0.904$, $H_{AB} = H + 1.374$ and $K_{AB}= K_s + 1.841$.

\begin{table}
\caption{Quality assessment of reduced images. }
\label{tab:qc}
\center
\begin{tabular}{lccccccc}
\hline\hline
Region & Passband &  A  &  B   & C  & D \\
\hline
CDF-S & $J$   &  13 & 0  &  0 & 0 \\
CDF-S & $H$   &  1  & 0  &  0 & 0 \\
CDF-S & $K_s$ &  10 & 3  &  0 & 0 \\
HDF-S & $J$   &  11 & 0  &  0 & 0 \\
HDF-S & $H$   &  7  & 2  &  0 & 0 \\
HDF-S & $K_s$ &  3  & 12 &  0 & 0 \\
\hline
\end{tabular}
\end{table}

Before creating stacks and associated catalogs, discussed in the next
section, all the reduced images were inspected and
graded. Table~\ref{tab:qc} summarizes the results of the visual
inspection grouping by mosaic. The table lists in Col.~1 the region
name; in Col.~2 the passband; and in Cols.~3--6 the grades given
ranging from A (best) to D (worst).  These grades reflect the overall
cosmetic quality of the images including the background and the
presence of other features that may affect the source extraction.  In
general, a grade A image shows no significant cosmetic problems, while
grade D indicates that this image is useless for any further
use. Grades B and C are intermediate, where C is usually not useful in
itself but can still add information in the subsequent stacking
procedure. In addition to the grade a subjective comment, to describe
any abnormality or feature, is normally associated with the image. As
can be seen from the table all images were graded A or B,
corresponding to a high cosmetic quality.

\begin{table*}
\caption{Summary of reductions.}
\label{tab:red}
\center
\begin{tabular}{lccccccc}
\hline\hline
Region & Passband & \# Fields & T (ksec) & Nights & Exposures & \#
Images \\
\hline
CDF-S  & $J$    & 4 & 40.1  & 3 & 668  & 13 \\
CDF-S & $H$     & 1 &  3.5  & 1 & 58   &  1 \\
CDF-S  & $K_s$  & 4 & 35.4  & 4 & 590  & 13  \\
HDF-S  & $J$    & 3 & 29.9  & 5 & 498  & 11  \\
HDF-S  & $H$    & 2 & 23.2  & 4 & 387  & 9 \\
HDF-S  & $K_s$  & 3 & 51.0  & 6 & 850  & 15 \\
\hline
\end{tabular}
\end{table*}

 The 3051 science frames were input to the EIS Data Reduction System
and converted into 62 reduced images, each corresponding to an
observing block, with 35 in the HDF-S and 27 in the CDF-S regions.
Table~\ref{tab:red} summarizes the information on the available
reduced images. The table gives in Col.~1 the region name; in Col.~2
the passband; in Col.~3 the number of fields observed; in Col.~ 4 the
total on-source time in ksec; in Col.~5 the number of nights in which
the observations were carried out; in Col.~6 the number of raw
exposures involved; in Col.~7 the number of reduced images available.

Typically, the reduction data rate is of the order of 0.2 Mpix/sec for
each dual processor. Since the system was operating with 8 independent
machines the effective data rate was of the order of 1.6 Mpix/sec. The
high-throughput of the system together with its administrative
infrastructure provided by its associated database and system
architecture was essential, allowing re-processing and re-calibration,
in particular, taking into account the large number of indivdual
programs carried out by the EIS team. This, together with the
infrastructure available for keeping different versions of products,
enabled the comparison of different reductions in an easy way, a key
element for long term programs.

\section {Final products}
\label{sec:finalproducts}

\begin{figure}
\center
\resizebox{0.7\columnwidth}{!}{\includegraphics{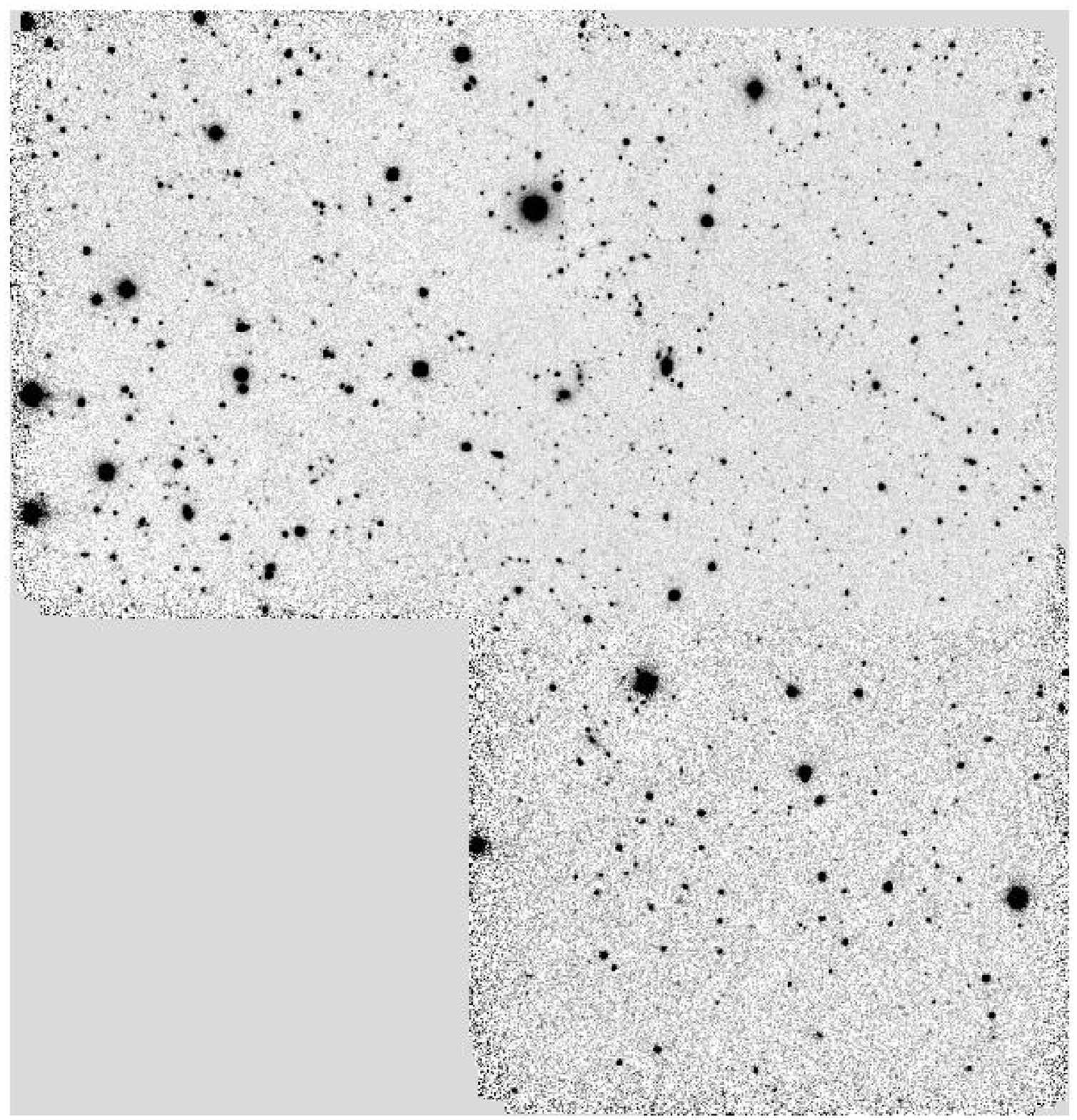}}
\resizebox{0.7\columnwidth}{!}{\includegraphics{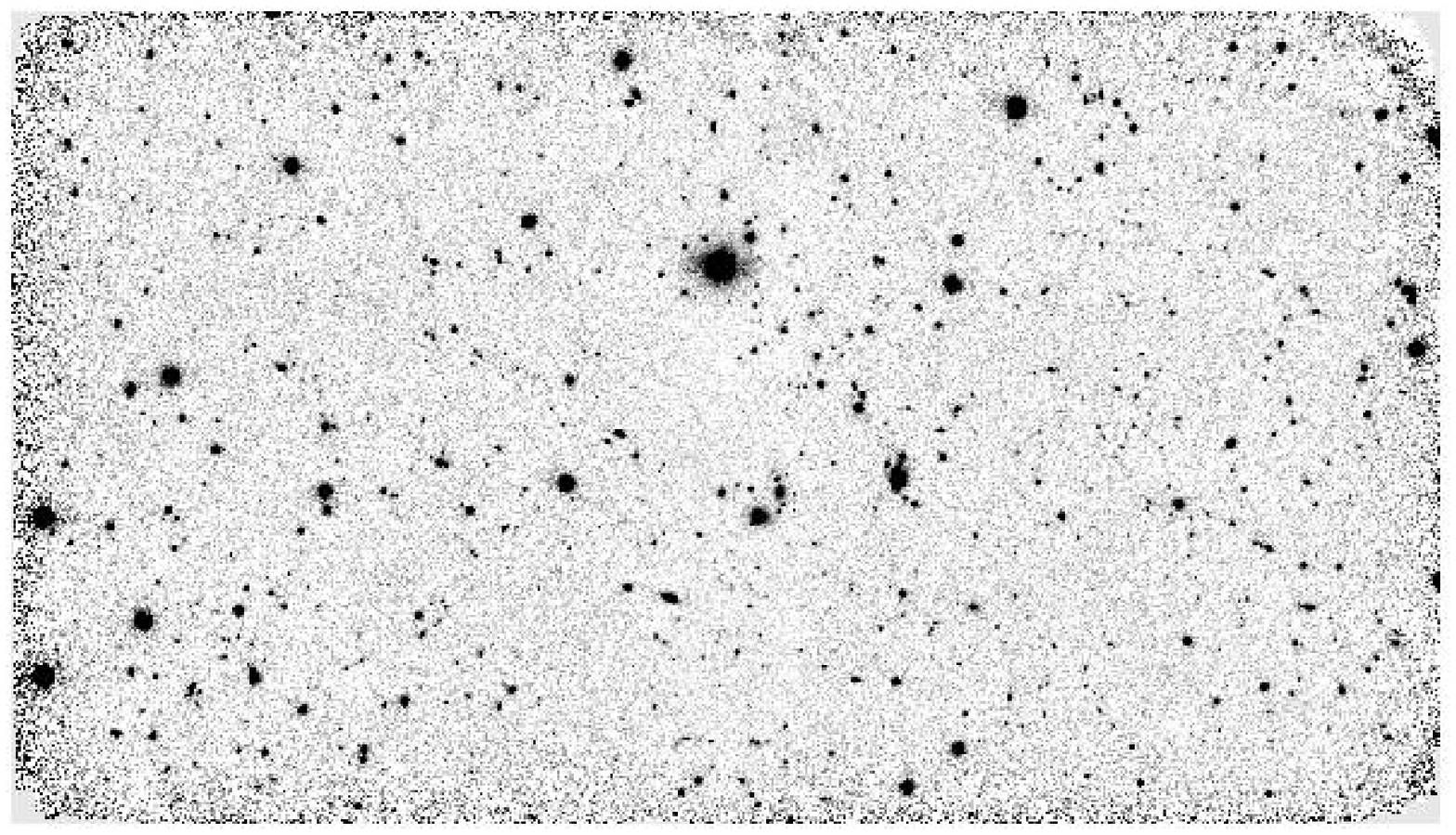}}
\resizebox{0.7\columnwidth}{!}{\includegraphics{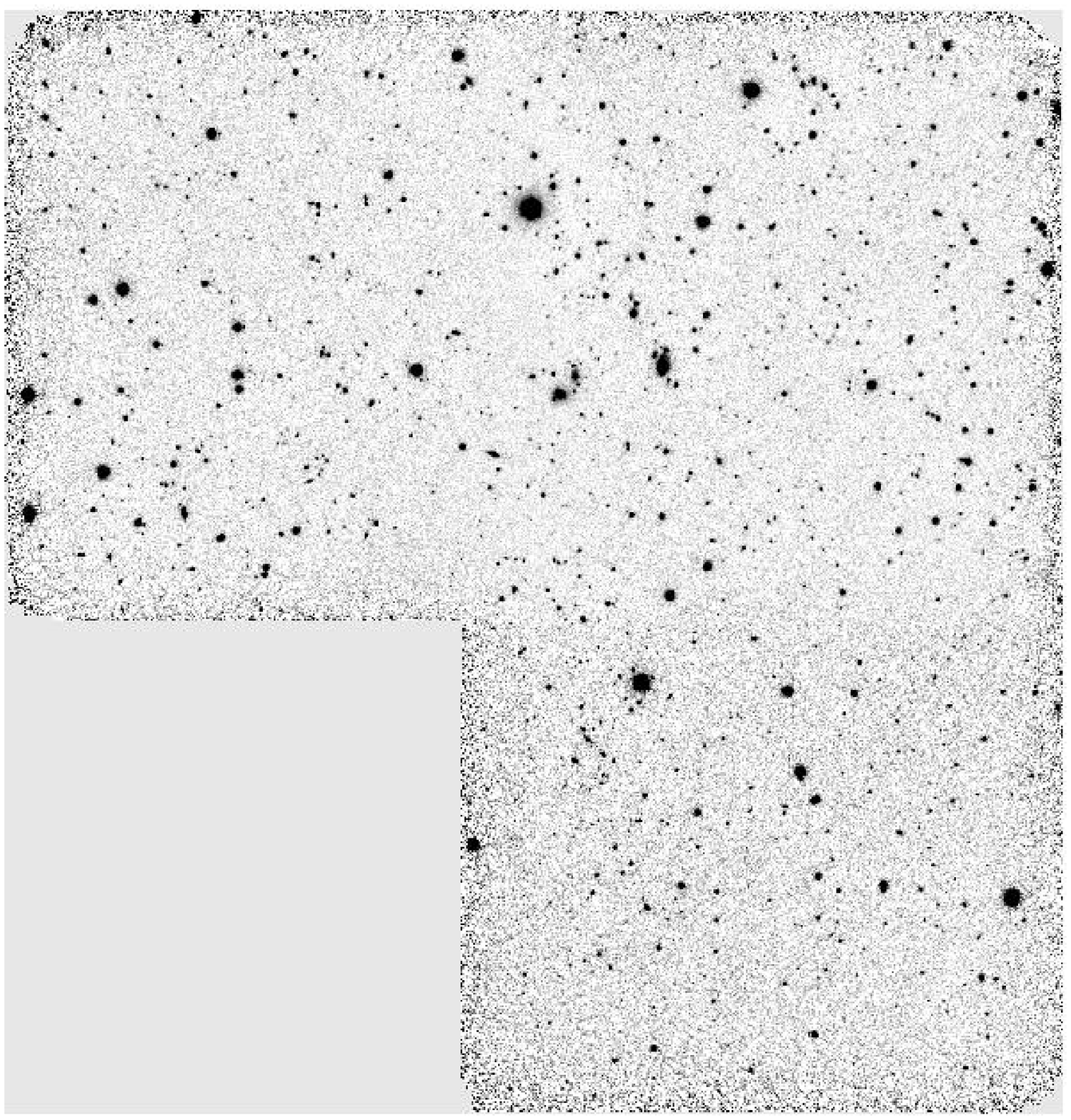}}
\caption{Mosaic image of the HDF-S in the \j , \h\ and \k -band from top to bottom.}
\label{fig:irmosaic_hdfs}
\end{figure}

\begin{figure}
\center
\resizebox{0.7\columnwidth}{!}{\includegraphics{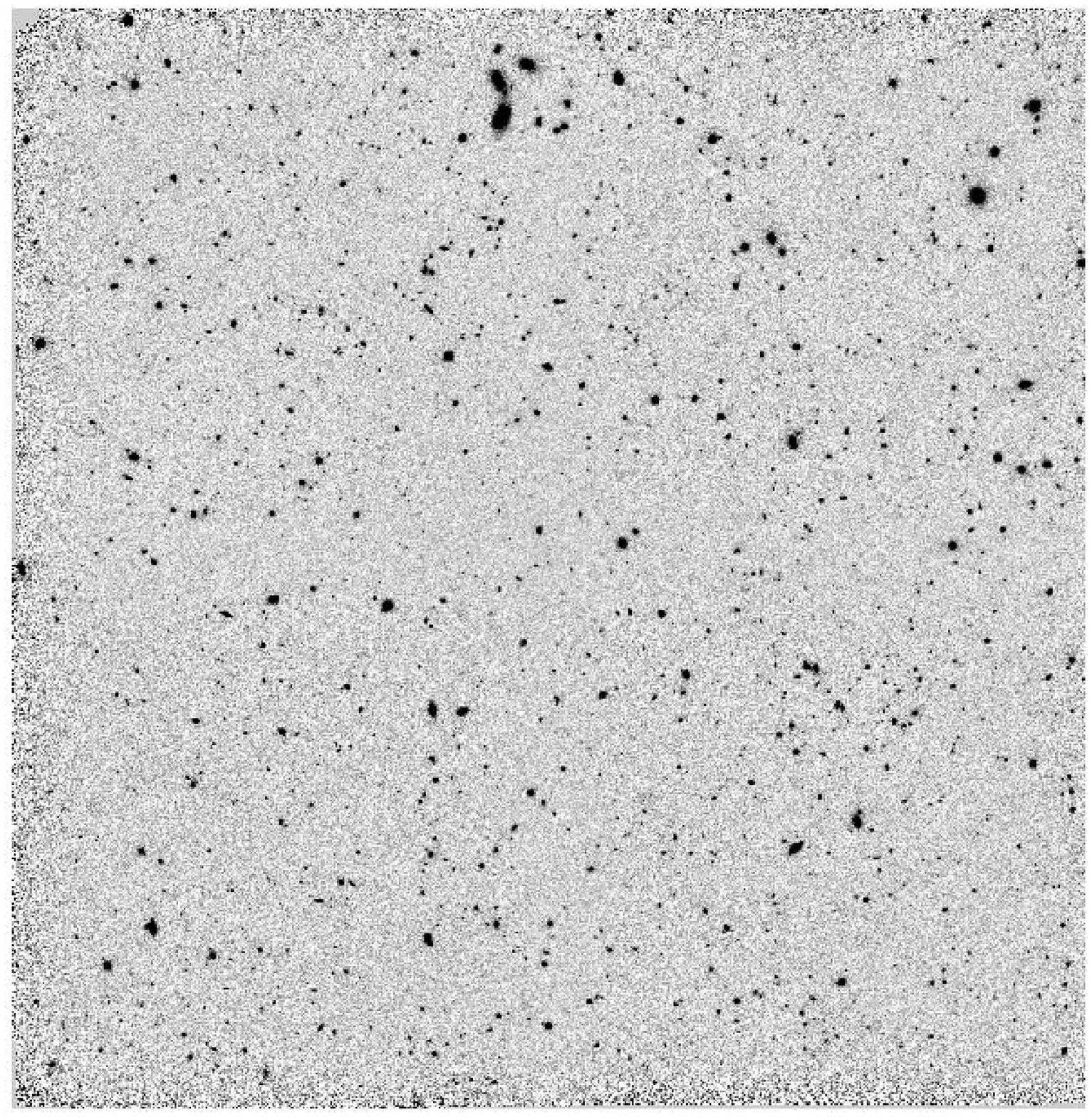}}
\resizebox{0.7\columnwidth}{!}{\includegraphics{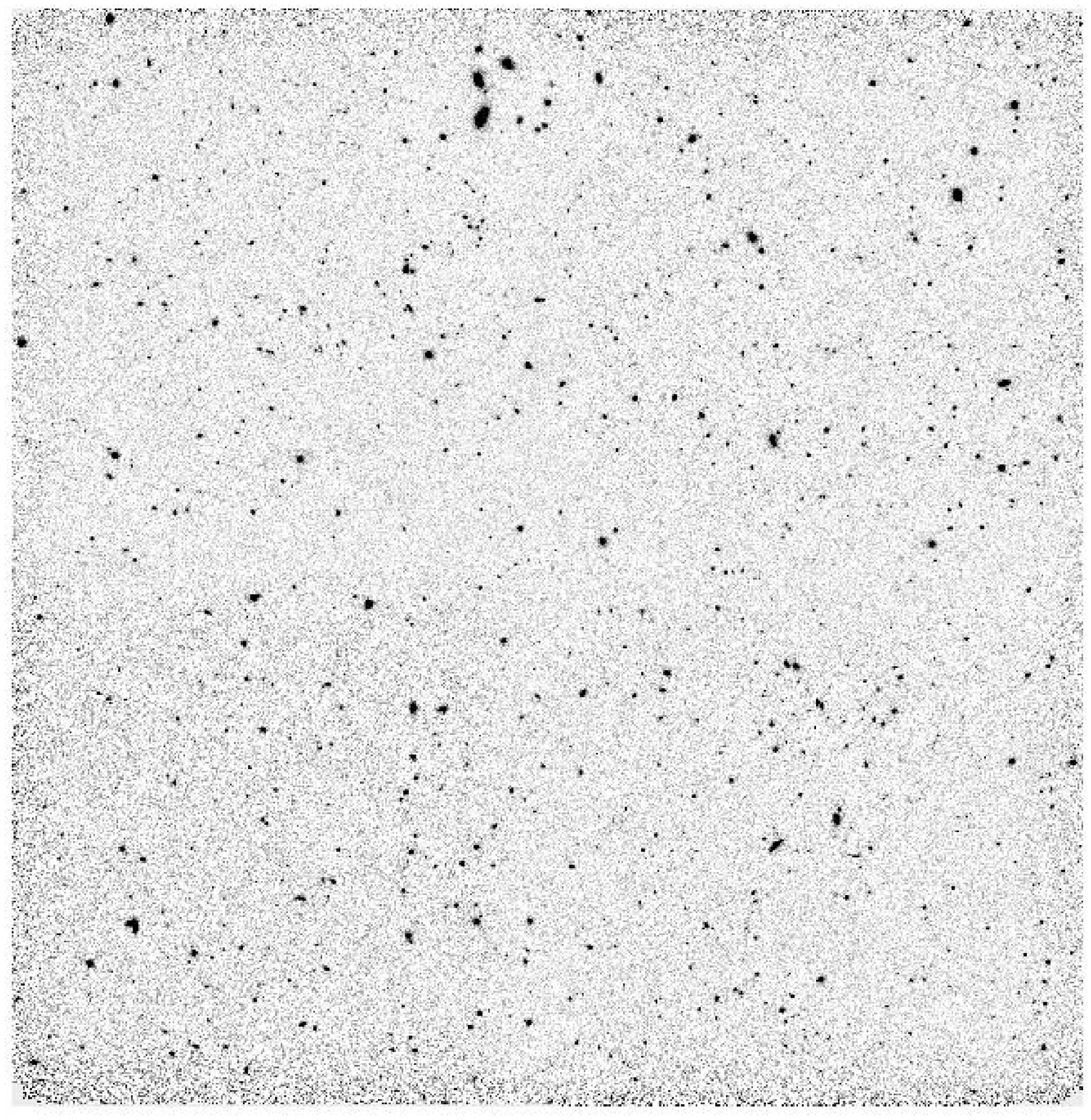}}
\caption{Mosaic image of the CDF-S in the \j\ (upper) and \k -band (lower).}
\label{fig:irmosaic_cdfs}
\end{figure}

\subsection {Images}

The 62 reduced images mentioned in the previous section were converted
into 17 stacked (co-added) images following the procedures described
in Paper~I. All stacked images were visually inspected and
graded. Again the quality was found to be good with 14 images being
graded A and the remaining 3 B.   The resulting mosaics in each
passband are shown in Figs.~\ref{fig:irmosaic_hdfs} and
\ref{fig:irmosaic_cdfs} for HDF-S and CDF-S, respectively.

\begin{table*}
\center
\caption{Main characteristics of the final stacks.}
\label{tab:attributes}
\begin{tabular}{llccrcccc}
\hline
\hline
Field & Passband & Grade & PSF FWHM & PSF rms & $m_{lim}$  (Vega) &
Completeness \\
\hline
CDF-S-1 & \j & A & 0.84 & 0.194 & 22.72 & 94\% \\
CDF-S-1 & \k & A & 0.97 & 0.183 & 21.08 & 105\% \\
CDF-S-2 & \j & A & 0.92 & 0.139 & 22.86 & 89\% \\
CDF-S-2 & \k & A & 0.70 & 0.189 & 20.90 & 90\% \\
CDF-S-3 & \j & A & 0.80 & 0.194 & 22.95 & 99\% \\
CDF-S-3 & \k & B & 1.12 & 0.123 & 20.75 & 84\% \\
CDF-S-4 & \j & A & 0.64 & 0.111 & 21.44 & 89\% \\
CDF-S-4 & \h & A & 0.84 & 0.113 & 21.21 & 32\% \\
CDF-S-4 & \k & A & 0.87 & 0.122 & 20.64 & 49\% \\
HDF-S-1 & \j & A & 1.22 & 0.113 & 22.79 & 156\% \\
HDF-S-1 & \h & B & 0.77 & 0.187 & 21.47 & 103\% \\
HDF-S-1 & \k & A & 0.82 & 0.212 & 21.34 & 208\% \\
HDF-S-2 & \j & A & 0.83 & 0.159 & 21.98 & 88\% \\
HDF-S-2 & \h & A & 0.84 & 0.181 & 21.69 & 112\% \\
HDF-S-2 & \k & A & 1.08 & 0.176 & 21.20 & 198\% \\
HDF-S-3 & \j & A & 1.17 & 0.060 & 22.07 & 33\% \\
HDF-S-3 & \k & B & 0.79 & 0.127 & 20.80 & 67\% \\
\hline
\end{tabular}
\end{table*}

The main attributes of the stacks produced for each mosaic are
summarized in Table~\ref{tab:attributes}. The table gives in Col.~1
the field name; in Col.~2 the passband; in Col.~3 the grade; in
Cols.~4 and 5 the point-spread function (PSF) FWHM in arcseconds and
its anisotropy (PSF rms) as measured in the final stacked image; in
Col.~6 the limiting magnitude, $m_{lim}$, estimated for the final
image stack for a 2\arcsec aperture, $5\sigma$ detection limit in the
Vega system; in Col.~7 the completeness of the observations expressed
as the fraction (in percentage) of observing time relative to that
originally planned. It can be seen that the seeing is, in general,
quite good, ranging from 0.64 to 1.22~arcsec with a median of
0.84~arcsec. The median limiting magnitudes in the Vega system are
22.72, 21.47 and 20.90 in the \j , \h , and \k -band, corresponding to
23.62, 22.84 and 22.74 in the AB-system.

As a final check of the quality of the photometric calibration,
the regions of overlap between adjacent frames were used to estimate
the relative field-to-field magnitude variations, yielding an estimate
of the overall uncertainty in the absolute calibration. In the HDF-S
there are only two overlaps in \j~ and \k, with the mean values being
0.05 and 0.07, and comparable  scatter. In the \h-band there is
only one overlap with an offset of 0.05. For the CDF-S there are four
overlaps for the \j~ and \k~data, yielding a mean offset of 0.03 and
0.02~mag and a scatter of 0.029 and 0.024~mag, respectively.  From
these results one can estimate the accuracy of the absolute
calibration of HDF-S is $\sim$ 0.05~mag, while that of CDF-S is
considerably better $\lesssim 0.03$~mag in both bands.

\subsection {Catalogs}
\label{cats}

\begin{table*}
\center
\caption{Main characteristics of the  catalogs extracted from the final
 images.}
\label{tab:catalogs}
\begin{tabular}{llccrcccc}
\hline
\hline
Field & Passband & Eff. area & \# Objects & $m_{lim}$ (Vega) \\
\hline
CDF-S-1 & \j & 26.3 & 448 & 21.56 \\
CDF-S-1 & \k & 25.9 & 362 & 19.56 \\
CDF-S-2 & \j & 25.6 & 453 & 21.78 \\
CDF-S-2 & \k & 25.2 & 320 & 19.67 \\
CDF-S-3 & \j & 26.3 & 459 & 21.45 \\
CDF-S-3 & \k & 26.3 & 352 & 19.76 \\
CDF-S-4 & \j & 25.9 & 431 & 21.68 \\
CDF-S-4 & \h & 25.9 & 227 & 19.95 \\
CDF-S-4 & \k & 25.9 & 230 & 19.13 \\
HDF-S-1 & \j & 25.9 & 460 & 21.59 \\
HDF-S-1 & \h & 25.9 & 442 & 19.64 \\
HDF-S-1 & \k & 25.9 & 496 & 19.30 \\
HDF-S-2 & \j & 26.3 & 672 & 21.87 \\
HDF-S-2 & \h & 25.6 & 564 & 19.81 \\
HDF-S-2 & \k & 26.3 & 521 & 20.10 \\
HDF-S-3 & \j & 25.2 & 262 & 20.81 \\
HDF-S-3 & \k & 26.3 & 395 & 19.34 \\
\hline
\end{tabular}
\end{table*}

Catalogs were extracted from the stacked images using the EIS Data
Reduction System adopting the same procedure as described in Paper~I,
which also describes the catalog format.  The final
science-grade catalogs presented include only objects with $S/N
\geq 5$ as computed from the magnitude error.
Table~\ref{tab:catalogs} summarizes the characteristics of the
extracted catalogs. The table gives in Col.~1 the field name; in
Col.~2 the passband; in Col.~3 the effective area in square arcmin; in
Col.~4 the number of objects; and in Col.~5 the 80\% completeness
limit in the Vega system, $m_{lim}$, as estimated from mock catalogs.

Each catalog and its associated product log, including extensive
comparison with other data previously ingested in the associated
database of the EIS system, takes about 2.5~min, corresponding to an
overall mean data rate of 0.03~Mpix/s.

\section{Discussion}
\label{discussion}

The EIS-DEEP infrared data presented here was used as the test case
for the implementation of both the stand-alone version of the image
processing library EIS/MVM, and the EIS Data Reduction System as a
whole. The latter is responsible for a variety of automatic and
un-supervised processes ranging from the creation of the reduction
blocks and of the reduced images on a nightly basis to the creation of
stacks and their associated final science-grade catalogs presented in
the previous section.  Because the EIS-DEEP infrared dataset is at the
same time relatively small, enabling extensive tests to be done, and
large enough to sample a variety of problems that emerge under real
observing conditions, thereby helping identify trouble-spots in the
code, it provides an excellent test case. Furthermore, this dataset
has been reduced both by the EIS team, using different software
packages \citep{dacosta98, rengelink98} or versions of the same
software \citep[][and the present paper]{vandame01}, and independently
by other groups using different techniques, thereby allowing the
comparison of different versions. Finally, several other datasets
covering the regions considered have been compiled since the
completion of the EIS-DEEP observations, using either the same or
different instruments with different limiting magnitudes. The
availability of these various datasets discussed below is in marked
contrast to the time these data were first released by the EIS team.

The availability of this large array of different datasets
covering the same region allows a careful assessment of the
performance of the EIS system and the quality of its final science
products by direct comparison of the position, flux scale and
absolute calibration of objects in common between source catalogs
extracted from different final images. Additionally,
statistical measures, here limited to number counts of galaxies
and stars, compared to those computed by other authors or
predicted by models are used to assess the overall consistency
with results from the literature.

\subsection{Comparison with other reductions}

The comparison with other reductions of the same dataset has been
extensively used to show, a posteriori, the shortcomings of previous
releases of this dataset either produced by the original
implementation of the program {\it Jitter} of {\it Eclipse}
\citep{devillard99} or by an earlier version of the new EIS/MVM image
processing library. Below, the same strategy is adopted to evaluate
the performance of the current version of the system also used in the
reduction of the infrared part of the DPS survey to be presented in a
forthcoming paper \citep{olsen06b} and of the optical data in
\cite{dietrich05, mignano06}.

\begin{figure}
\center
\resizebox{0.7\columnwidth}{!}{\includegraphics{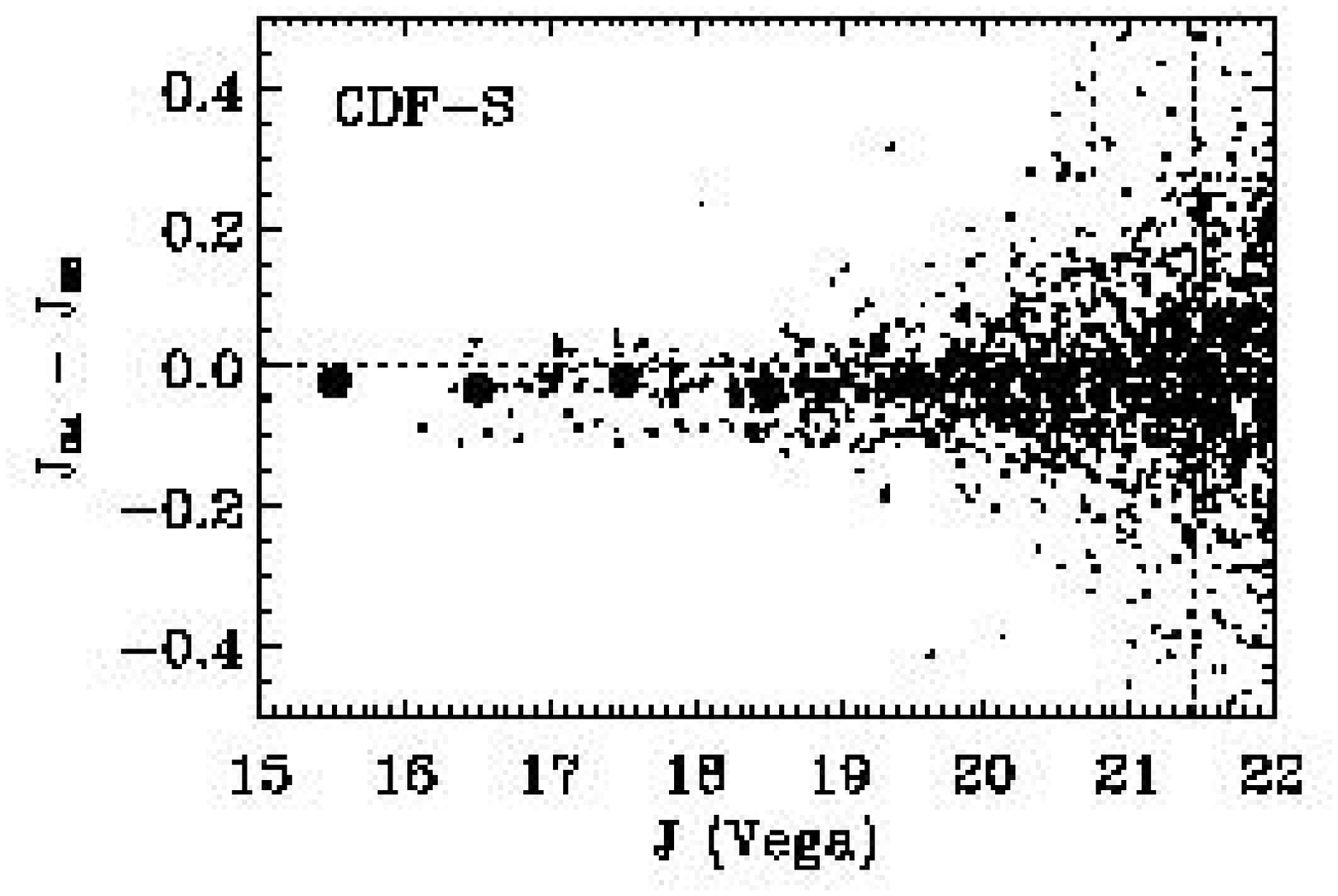}}
\resizebox{0.7\columnwidth}{!}{\includegraphics{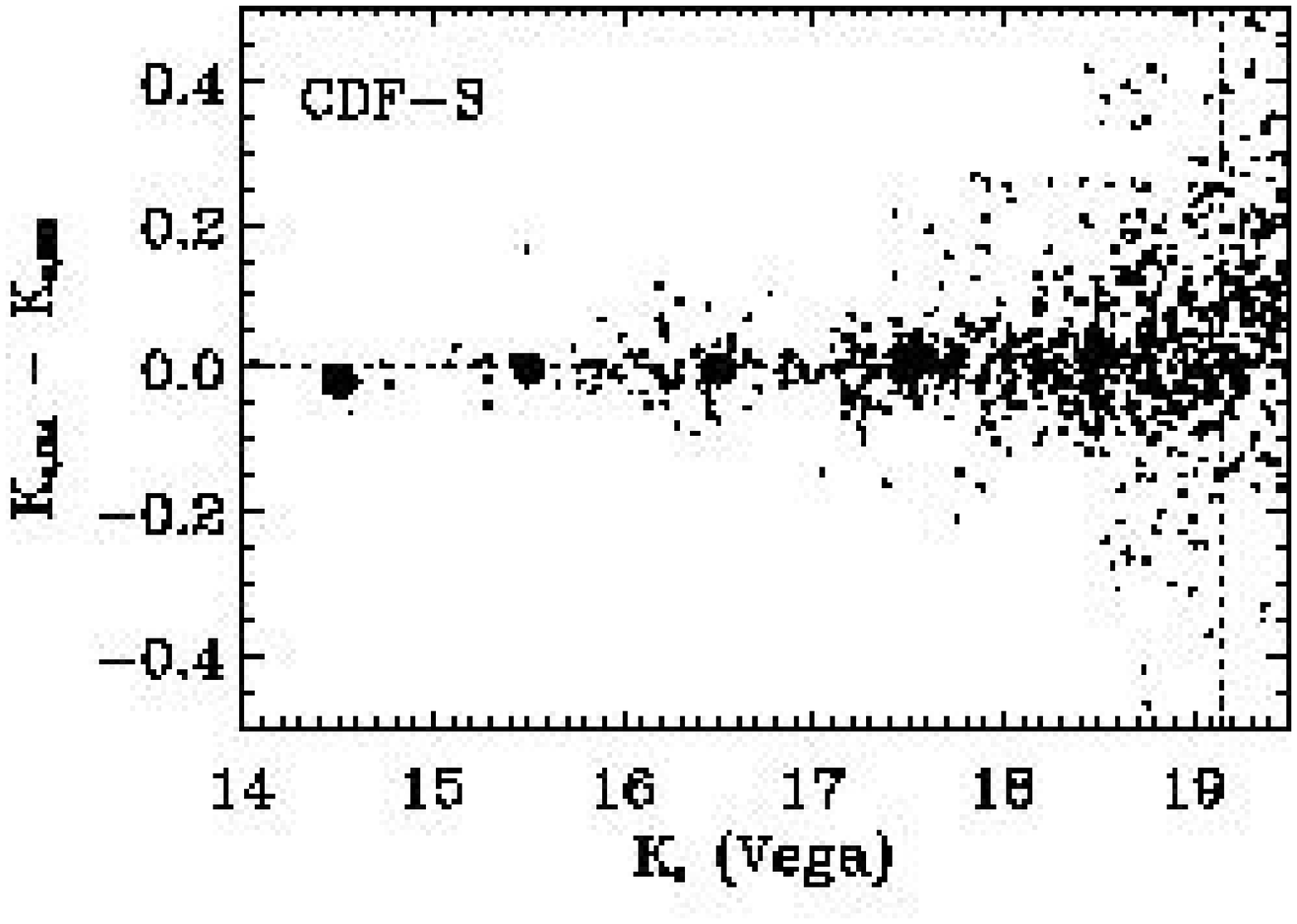}}
\caption{Comparison between the measured $J$ (upper panel),
 and $K_s$ (lower panel) magnitudes of objects in common with those
from \cite{vandame01}. The large filled circles represent
the mean magnitude difference computed in bins of 1~mag, while the
error bars denote the scatter in each bin. All magnitudes are in
the Vega system.}
\label{fig:phot_vandame}
\end{figure}

As a starting point, Fig.~\ref{fig:phot_vandame} shows the difference
in magnitude of objects identified in images produced using an earlier
version of the EIS/MVM library, as reported in \cite{vandame01}, and
those of the most recent version of the package, in all four fields of
the CDF-S region. The objects were matched by positional coincidence,
using a search radius of 1~arcsec, and only 5$\sigma$ detections are
considered. These values are used throughout this paper.  The large
filled circles are the mean value of the magnitude differences in bins
of one magnitude, and the error bars indicate their scatter. The
vertical dashed lines in the figure denote the brightest magnitude
$m_{lim}$, of the four fields considered, where $m_{lim}$ is the
magnitude at which the completeness of the catalog drops to 80\%.
From the top panel one finds that the error in the earlier version of
the image processing library had little effect in the relative
photometry of the \j -band data, with the possible exception of the
faintest magnitude bin, which shows the hallmark of lost flux in the
older reduction. Surprisingly, after re-calibration one also finds
that the \j~magnitudes for one of the fields exhibits an offset with
respect to the others of as much as 0.1~mag. Closer inspection of the
data shows that this is associated to the field CDF-S-4. From the
bottom panel, one finds that in the \k-band the effects of flux loss
are slightly larger but still of the order of $\lesssim$ 0.05~mag,
significantly smaller than the originally reported bias, which reached
0.15~mag at the faint end.

\begin{figure*}
\begin{center}
\resizebox{0.45\textwidth}{!}{\includegraphics{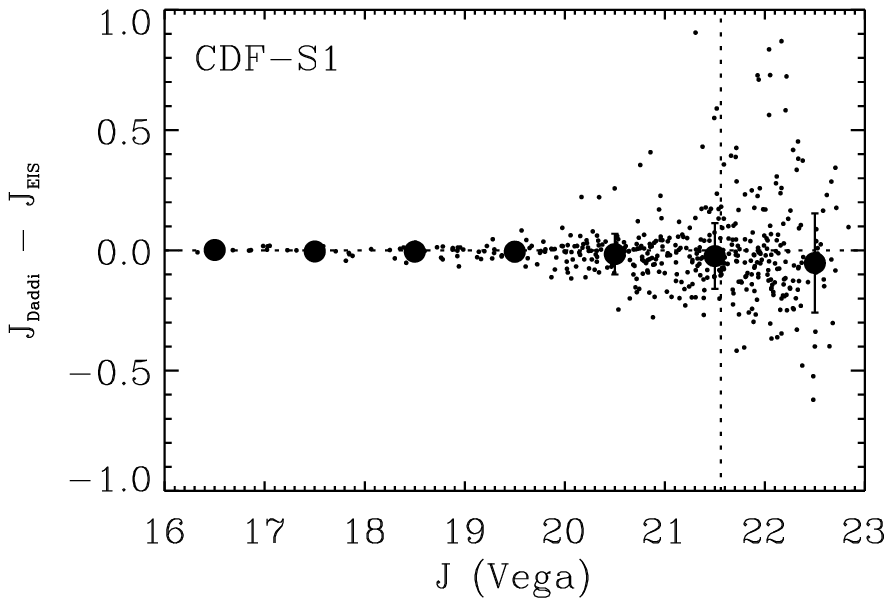}}
\resizebox{0.45\textwidth}{!}{\includegraphics{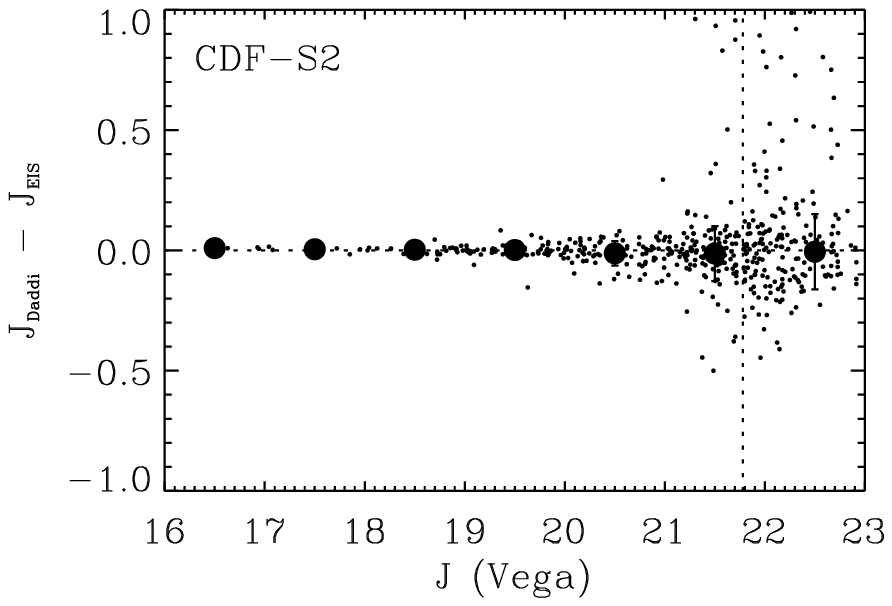}}
\resizebox{0.45\textwidth}{!}{\includegraphics{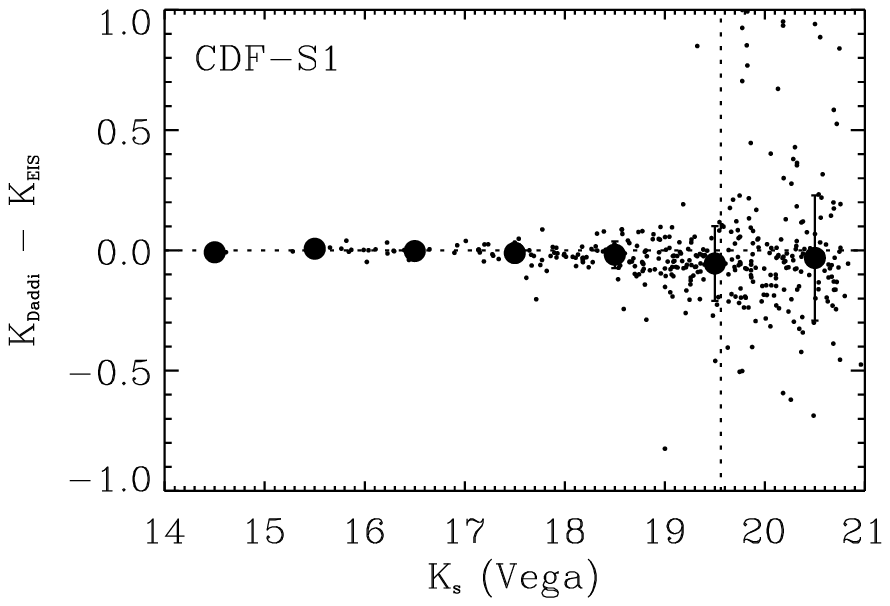}}
\resizebox{0.45\textwidth}{!}{\includegraphics{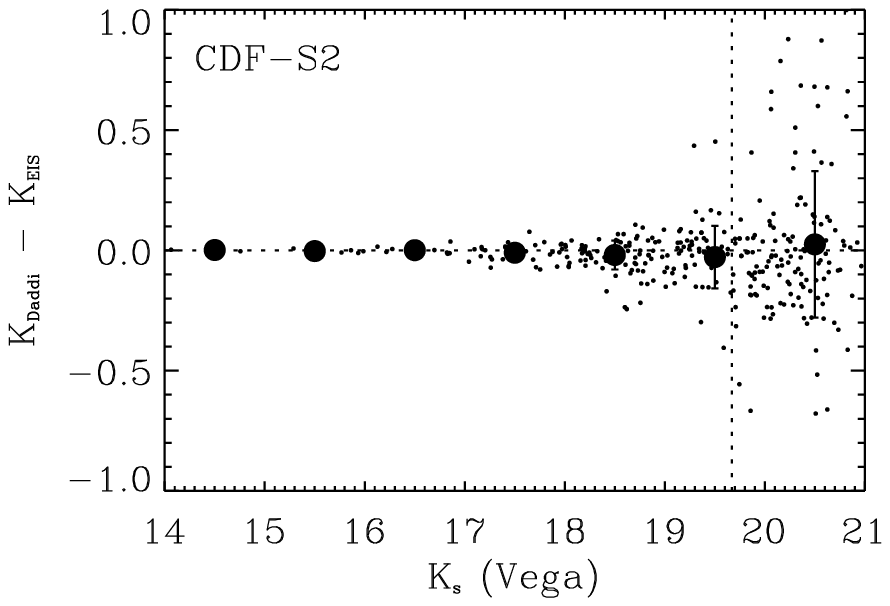}}
\caption{Difference in magnitudes \j\ (top panels) and \k\
(lower panels) as measured reducing the infrared data using the
DIMSUM package (Daddi private communication) and the EIS software.
As before, the large filled circles are the mean values of the
difference in bins of one magnitude and the error bars are the
scatter.}
\label{fig:photo_daddi}
\end{center}
\end{figure*}

To further verify the quality of the produced stacks, the
\j - and \k -band images of the CDF-S fields 1 and 2  reduced using the IRAF
package DIMSUM (Daddi, private communication) were used to extract
catalogs, in the same way as those created by the EIS system. The
catalogs were matched and the magnitudes of the objects in common
compared. In this comparison, the absolute calibration has been
neglected and any residual offset at the bright end between the two
sets of catalogs was arbitrarily removed. The results are shown in
Fig.~\ref{fig:photo_daddi} with the top panels showing the results for
the \j-band and the bottom panels for the \k-band, for the fields
indicated in each panel.   As can be seen from the comparison in both
bands and fields the results are excellent, demonstrating that the
earlier problems identified in the image processing system have been
properly addressed.

Similar results were obtained comparing other reductions of SOFI and
ISAAC data, reduced using either the same or different software
(e.g. Saracco, private communication; Dickinson, private
communication).  In summary, it has been shown that the current
reductions of the infrared data obtained using the EIS Data Reduction
System is free from any flux bias at faint magnitudes, yielding
results in excellent agreement with those obtained using other
software packages designed specifically to deal with infrared data.

\subsection {Comparison with other datasets}

\begin{figure*}
\begin{center}
\includegraphics[width=0.42\textwidth]{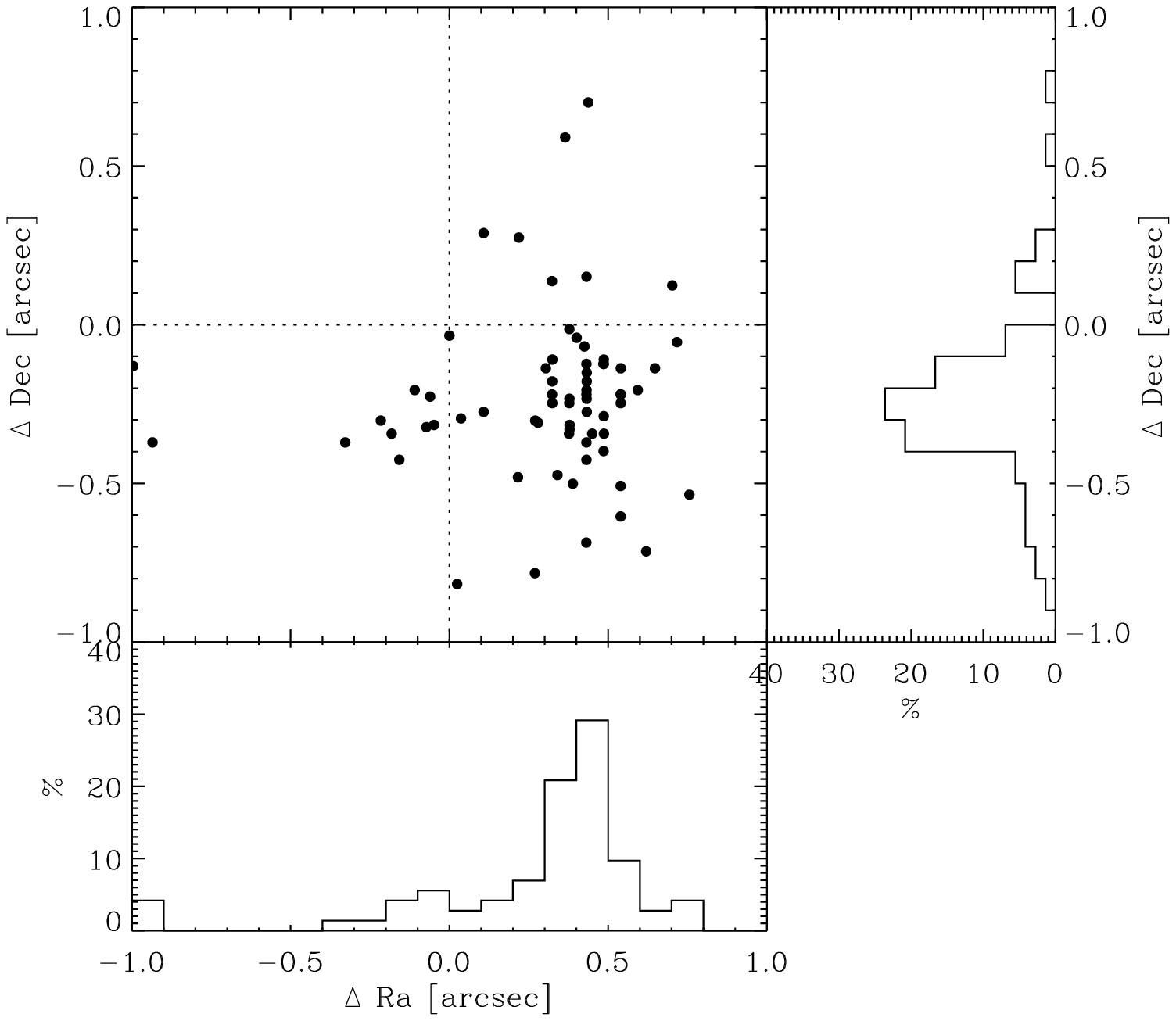}
\includegraphics[width=0.42\textwidth]{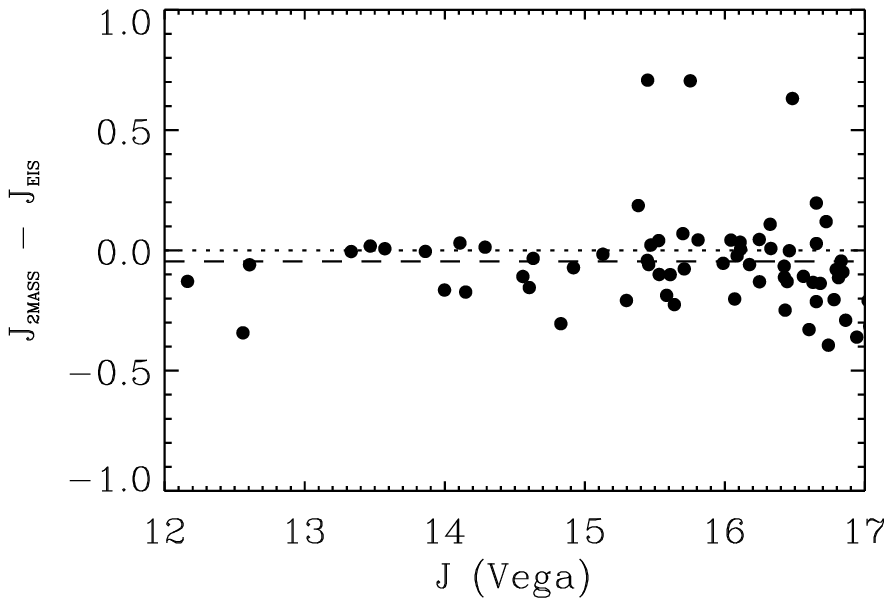}
\includegraphics[width=0.42\textwidth]{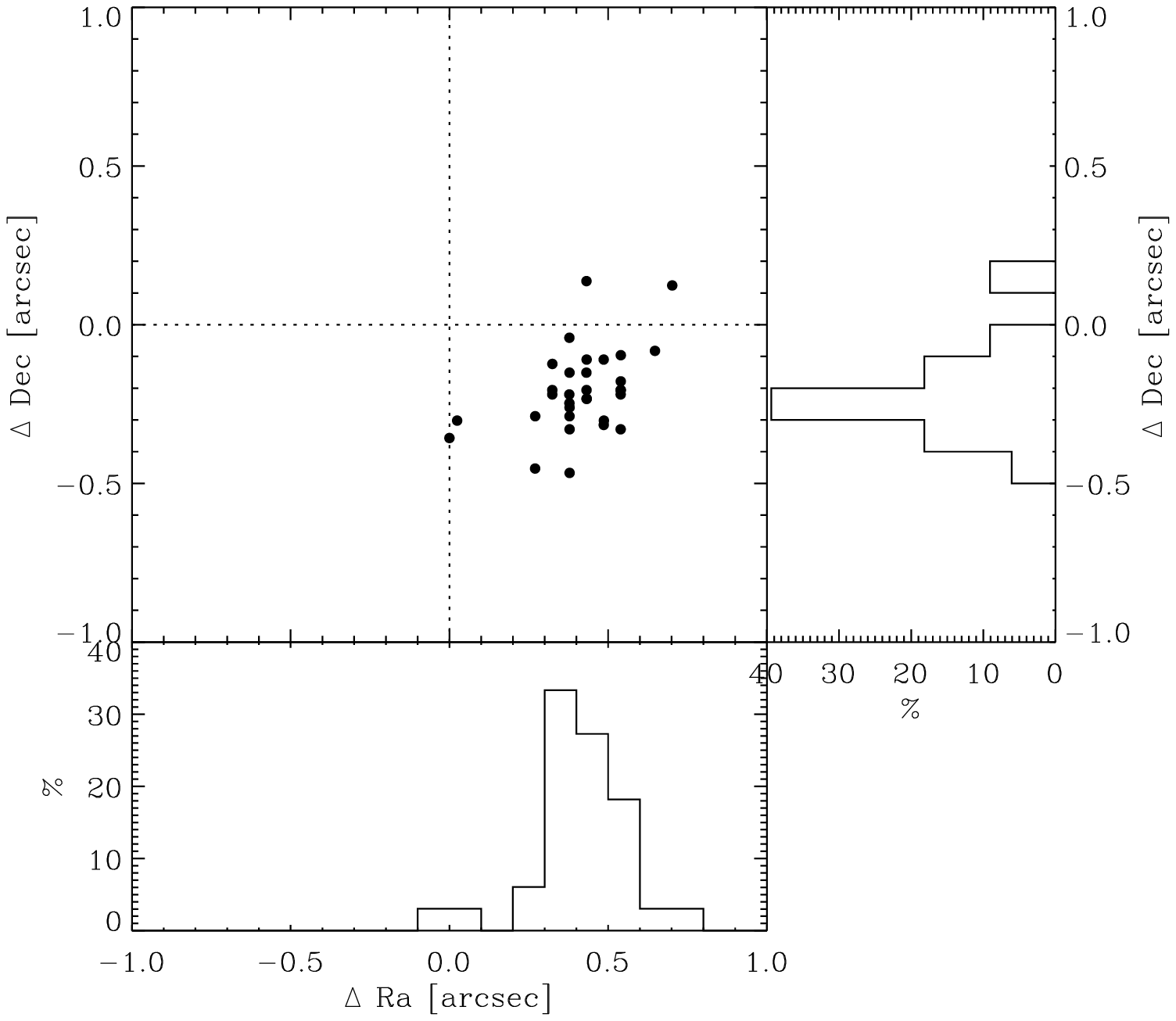}
\includegraphics[width=0.42\textwidth]{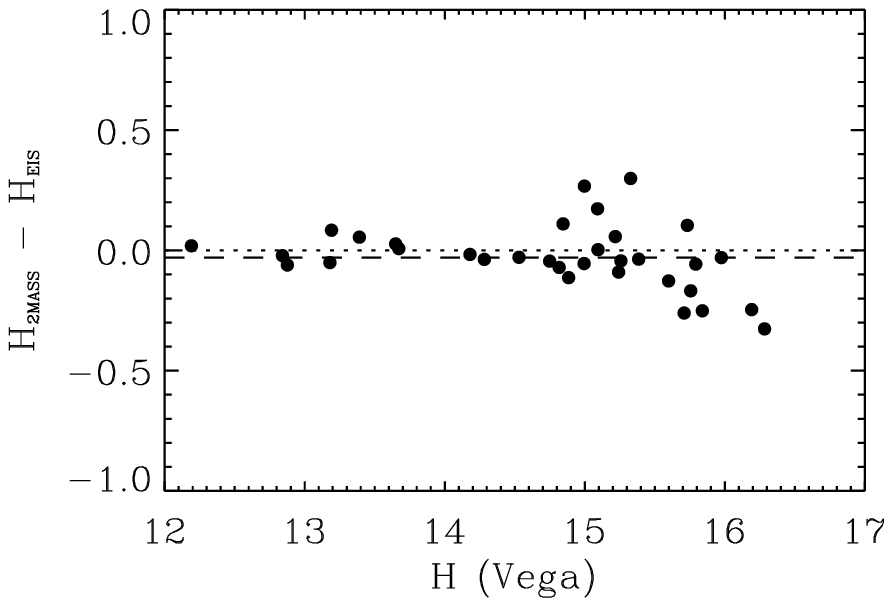}
\includegraphics[width=0.42\textwidth]{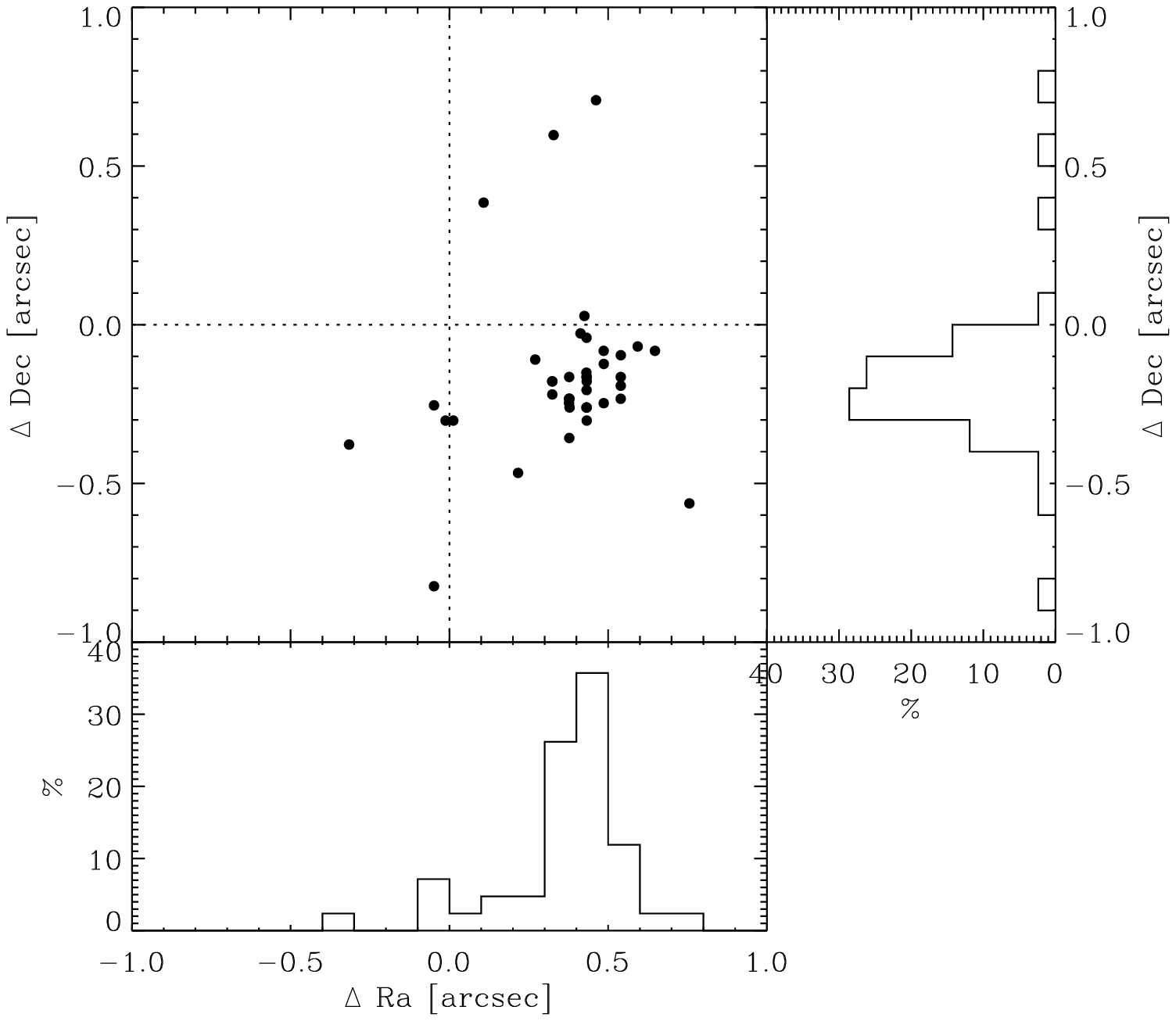}
\includegraphics[width=0.42\textwidth]{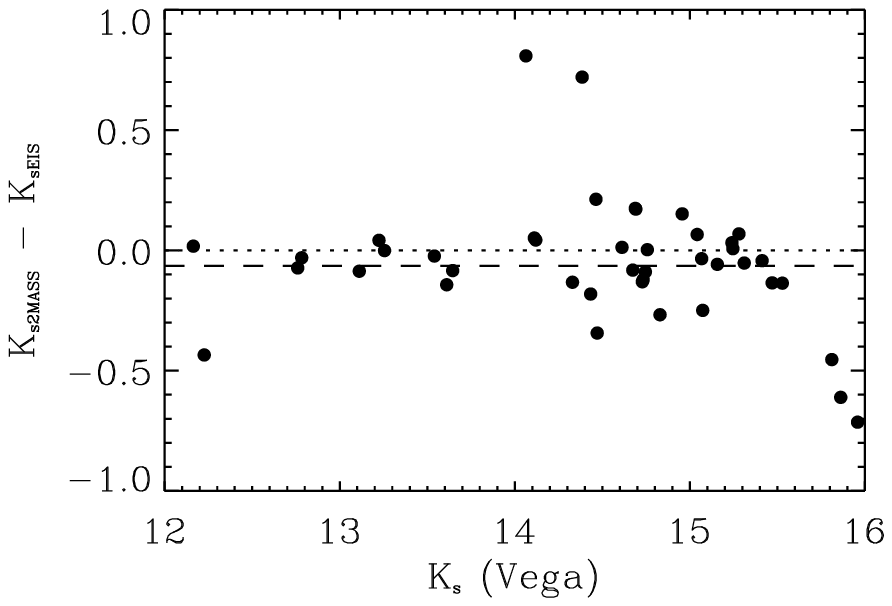}
\caption{Position (left panels) and magnitude (right
panels) differences (2MASS-EIS) between EIS sources in common with
those identified by the 2MASS survey \citep{kleinmann94} showing
the results in different passbands with \j\  on top and \k\ on the
bottom.}
\label{2mass}
\end{center}
\end{figure*}

\begin{figure*}
\center
\resizebox{0.32\textwidth}{!}{\includegraphics{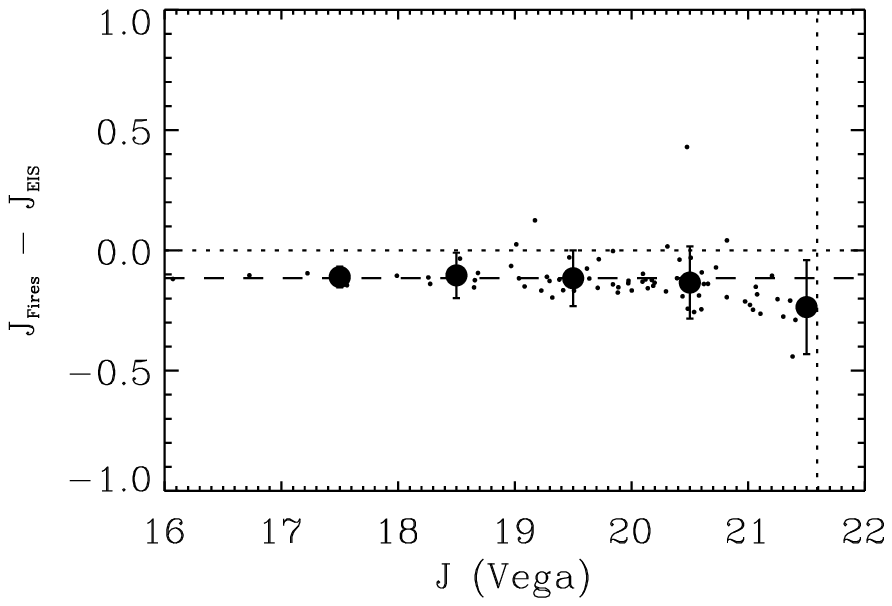}}
\resizebox{0.32\textwidth}{!}{\includegraphics{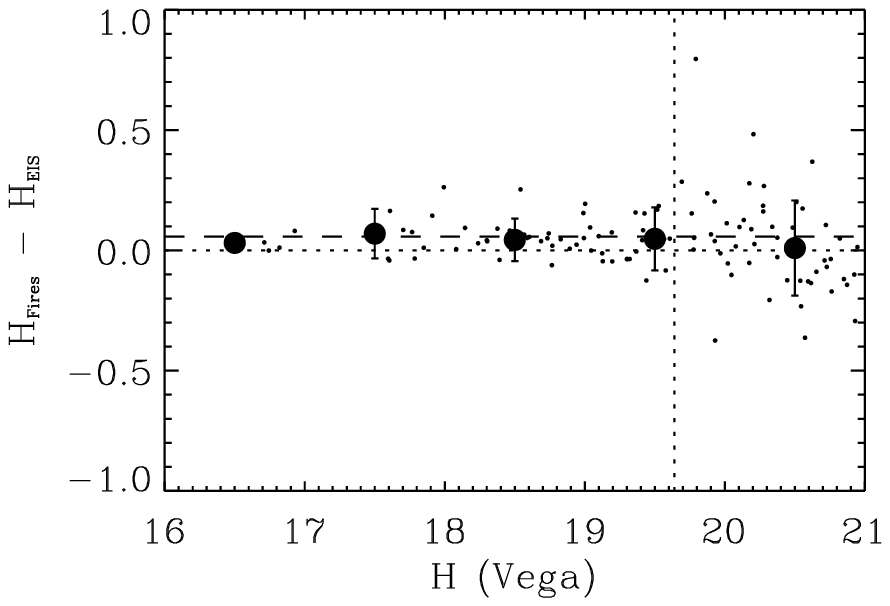}}
\resizebox{0.32\textwidth}{!}{\includegraphics{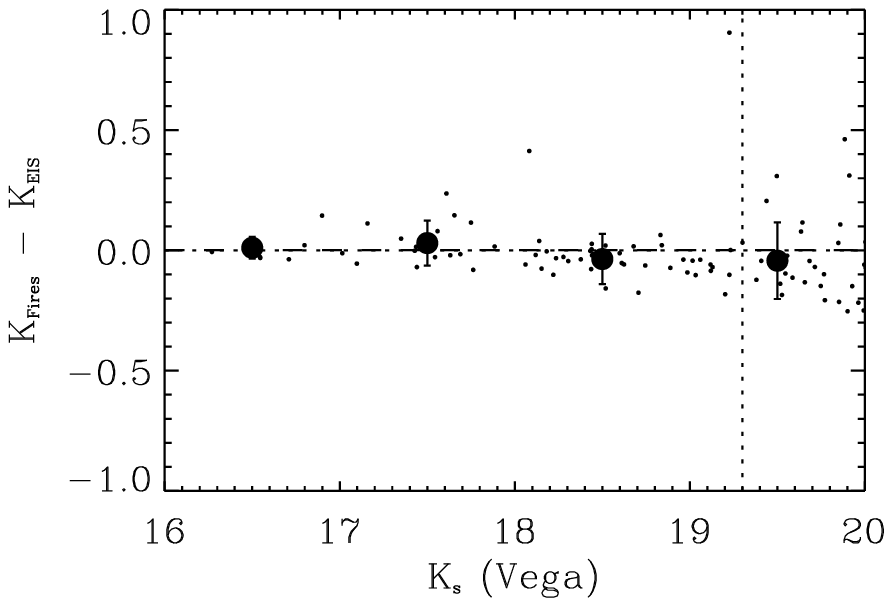}}
\caption{Magnitude difference (FIRES-EIS) between EIS
sources in common with FIRES \citep{labbe03} objects, as a
function of the EIS magnitude. The filled circles and error bars
are defined as in the previous figure. All magnitudes are in the
Vega system.} 
\label{fires}
\end{figure*}

\begin{figure*}
\center
\resizebox{0.95\columnwidth}{!}{\includegraphics{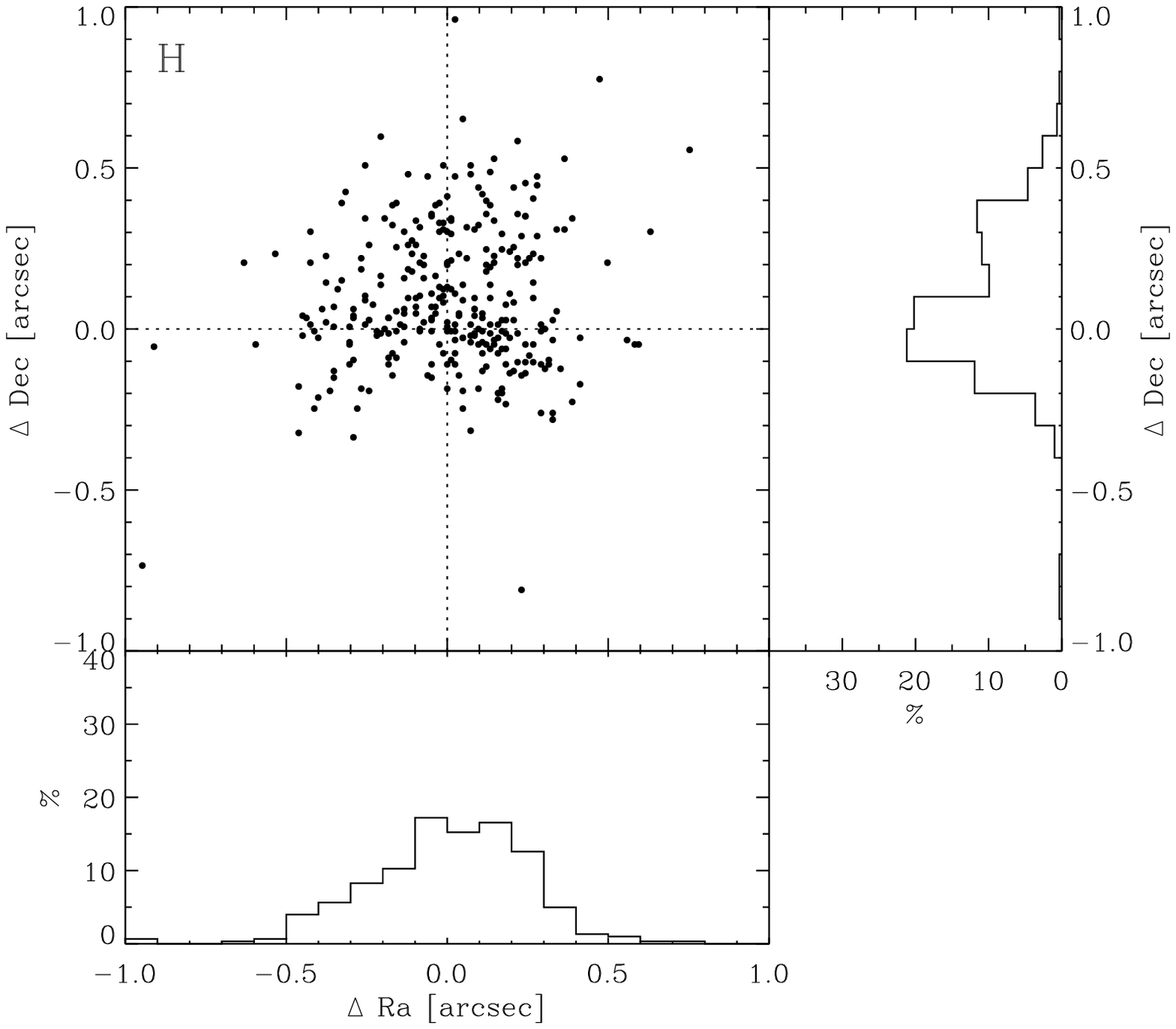}}
\resizebox{0.95\columnwidth}{!}{\includegraphics{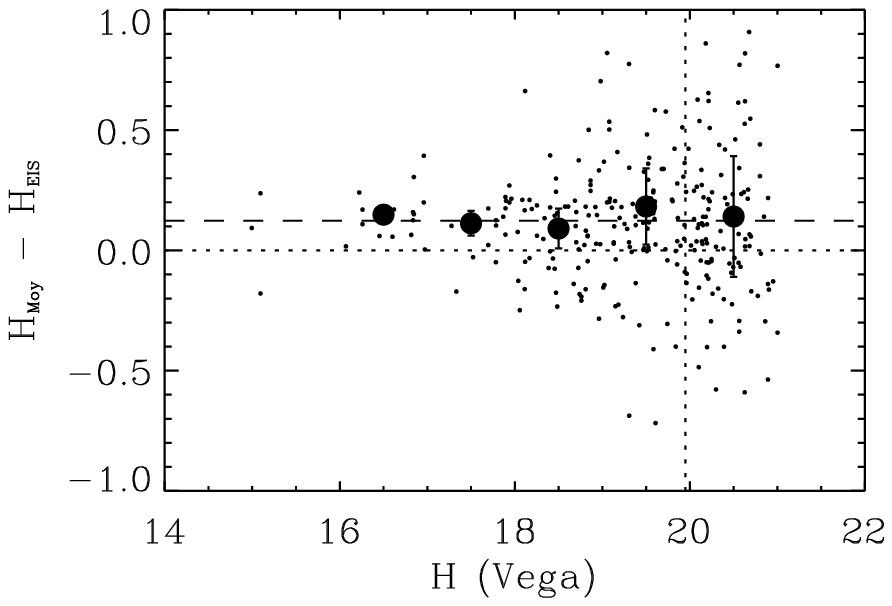}}
\caption{Position (left panel) and magnitude differences
(right panel) between EIS sources in common with those identified
by \cite{moy03} in the CDF-S region, as a function of the EIS
magnitude. The filled circles and error bars are defined as in
the previous figure. All magnitudes are in the Vega system.}
\label{moy}
\end{figure*}

\begin{figure*}
\center
\resizebox{0.95\columnwidth}{!}{\includegraphics{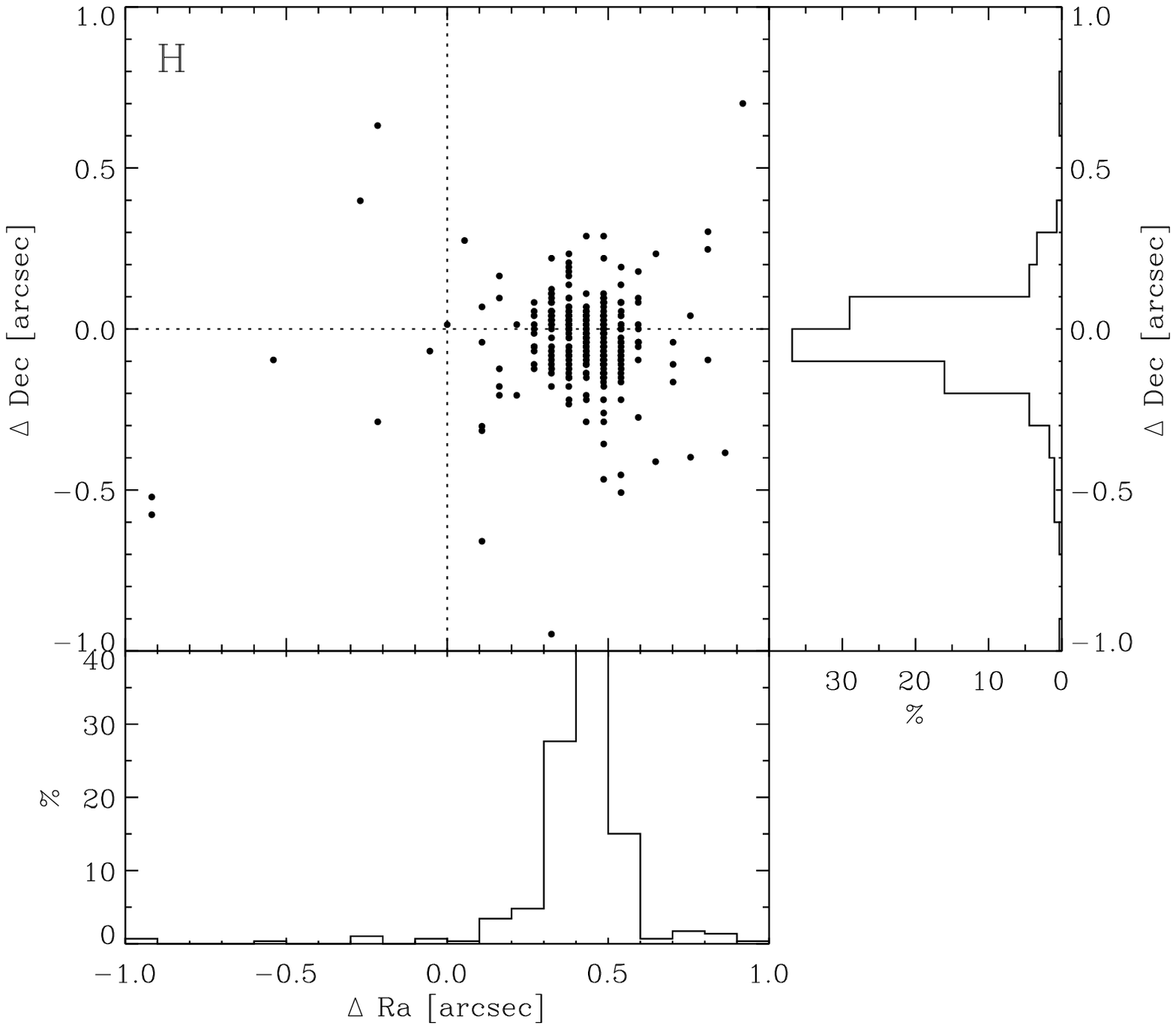}}
\resizebox{0.95\columnwidth}{!}{\includegraphics{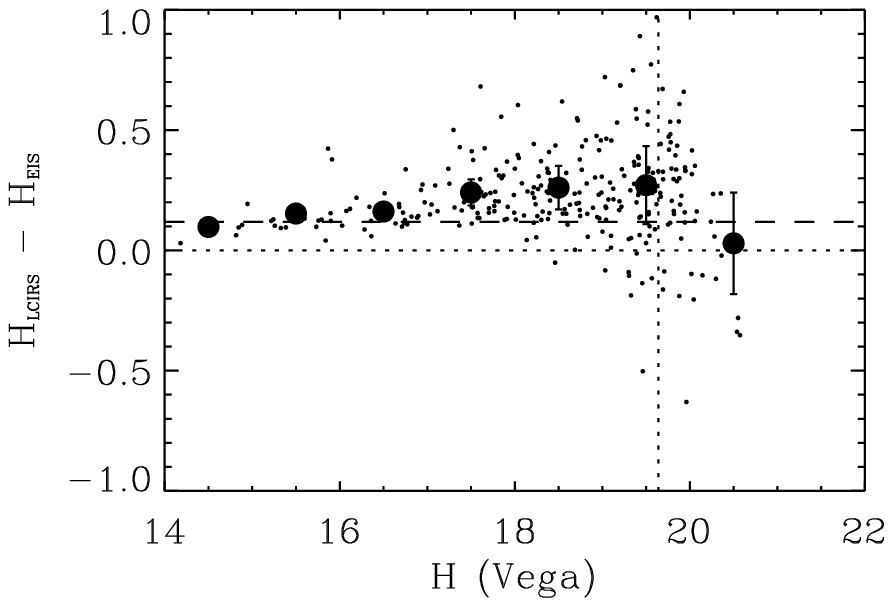}}
\caption{Same as Fig.~\ref{moy} but now comparing
positions and magnitudes measured by the LCIRS of the HDF-S.}
\label{lcirs}
\end{figure*}

The validation of the infrared reductions presented here
can be extended, comparing them to other fully calibrated datasets
available in the literature and covering the same regions surveyed
by EIS-DEEP. Examples of these datasets include: 1) data from the
Two Mass All Sky Survey \citep[2MASS, ][]{kleinmann94} in all
three passbands, useful to check the bright end; 2) deep data
taken with ISAAC at the VLT for the FIRES survey \citep{labbe03}
covering one of the pointings of the HDF-S region; 3) data taken
in H-band of the CDF-S region by \cite{moy03}; and data from the
Las Campanas Infrared Survey \citep[LCIRS, ][]{chen02} of the HDF-S region. 

The results of these various comparisons are shown in
Figs.~\ref{2mass}--\ref{lcirs}.  Fig.~\ref{2mass} shows the offset in
position (left panels) and magnitude (right panels) for objects in
common with 2MASS within 1~arcsec of EIS-DEEP detections. Each row
shows the results obtained combining all the fields for the different
bands with \j\ on top and \k\ on the bottom. While the number of
objects is relatively small, considering the magnitude interval of the
two samples, the main advantage of 2MASS is that data are available
for all fields allowing the use of a single reference to check both
the astrometric and photometric calibration. From Fig.~\ref{2mass} one
finds that there is a relative offset between the reference system of
the two datasets. However, this offset is nearly the same for all
passbands ranging from 0.27 to 0.43 arcsec in right ascension and from
0.19 to 0.26 arcsec in declination with a scatter of $\lesssim
0.40$~arcsec in all bands. Since the objects in common are
predominantly from HDF-S, the observed offset reflects the difference
between the Tycho and GSC2 coordinates in the HDF-S region. As shown
by \citep{olsen06b}, no systematic shifts are found relative to the
USNO-B catalog used to calibrate the different regions covered by the
DPS infrared dataset.

Comparison of the magnitudes yields offsets of -0.02, 0.004, and
-0.05~mag in \j, \h\ and \k, respectively, if one discards very bright
objects which tend to be partly saturated in the deep data. 
  Another important fact is that, except for the Malmquist bias seen
in the plots, there is no evidence of systematic trends with magnitude
in the interval considered. Furthermore, the small scatter suggests
some degree of consistency between the calibration of the images
involving as much as 7 fields distributed over two distant regions of
the sky, in some cases observed in different nights. All this
evidence suggests a consistent absolute calibration of the EIS-DEEP
data.

The behavior at faint magnitudes is evaluated by comparing the
present data to the deep observations of the FIRES survey
\citep{labbe03}. Since the authors do not provide the data with an
astrometric solution a crude shift has been applied to obtain a
proper match between the final images derived for the two
datasets. Therefore, in this case only magnitudes are
compared. Furthermore, in the comparison only objects detected at
5$\sigma$ in both catalogs are considered. Fig.~\ref{fires} shows
the result of this comparison for all the available passbands.
From the figure one finds that regardless of the passband, there
is no systematic deviation of the magnitude differences as a
function of the magnitude apart from a constant offset. One
possible exception is the difference at the faintest bin in the
\j-band, which seems to reflect the effects of a Malmquist bias
rather than flux-loss at the faint end. The comparison yields
offsets of -0.11, 0.06 and 0.001~mag in \j, \h\ and \k,
respectively.

In the \h-band two more datasets are available for comparison.  One is
that of \cite{moy03} covering the CDF-S region, for which there is an
overlap with the somewhat shallow EIS-DEEP observations of CDF-S-4,
and the other is that obtained as part of the LCIRS coverage of the
HDF-S region. The results of the comparison with \cite{moy03} are
shown in Fig.~\ref{moy}, where the differences in position (left
panel) and magnitude (right panel) of objects in common are
plotted. The mean offset in position is less than 0.1~arcsec with an
rms of about 0.45~arcsec, consistent with both having an error of the
order of $\sim0.25$~arcsec.  This result shows that the positions in
the CDF-S region have no systematic shift. The magnitudes are also in
good agreement except for an offset of 0.1~mag, somewhat larger than
the 0.06~mag derived from the comparison with FIRES data of HDF-S. The
origin of this offset is unknown and may be due to an inadequate
absolute calibration. However, this calibration can only be improved
by additional observations under photometric conditions and a better
calibration plan.  Again there is no systematic deviation of the
magnitude differences over the entire magnitude interval considered.
This is in marked contrast with the results obtained comparing the EIS
data with the LCIRS \h-band data taken in the region of the HDF-S,
shown in Fig.~\ref{lcirs}. Here the magnitude difference tends to
increase for fainter objects, with the LCIRS magnitudes being fainter
than those assigned by the EIS system. This trend reaches about
0.2~mag relative to the bright end, except for the faintest bin shown
affected by the Malmquist bias. Since this effect is not seen in any
of the other comparisons discussed above, it suggests that this effect
stems from the LCIRS reductions. Note that the offset in the bright
end is $0.12\pm0.07$, consistent with that computed from the
comparison with \cite{moy03} of 0.1~mag, but somewhat larger than the
offset of 0.06~mag measured against FIRES. Still, regardless of the
exact value this result also suggests a remarkable consistency in the
photometric calibration of the CDF-S and HDF-S regions.  On the other
hand, the left panel of Fig.~\ref{lcirs} shows the same offset in the
coordinates as observed relative to the 2MASS data. Note that the
absolute differences can easily be corrected. To construct
optical/infrared catalogs the astrometric calibration of infrared
images should be done using catalogs extracted from the optical
images, thereby minimizing possible relative offsets.

\subsection{Comparison of galaxy and star counts}

It is not always possible to assess the quality of the data by direct
comparison with overlapping observations. An alternative is to compute
different statistical measures, the simplest examples being galaxy and
star counts, and compare them with those obtained by other authors in
other regions of the sky, typically used for galaxies, or for stars
with those predicted by models of the Galaxy.

\begin{figure}
\begin{center}
\resizebox{\columnwidth}{!}{\includegraphics{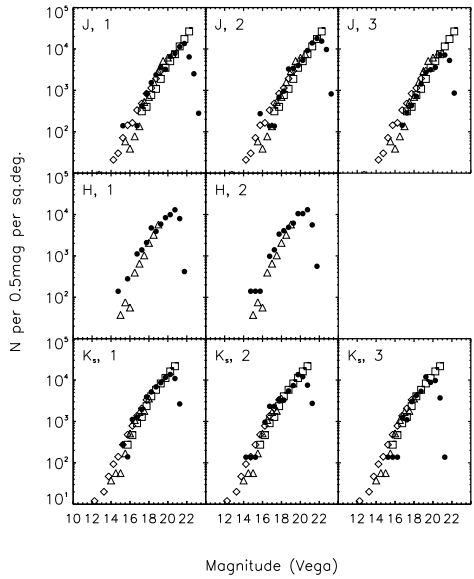}}
\caption{Galaxy number counts computed from the present data
 for each band (rows) and field (columns) of the HDF-S region (solid
 circles) compared to the published values with diamonds denoting
 those of \cite{vaisanen00}; triangles \cite{martini01} and squares
 \cite{iovino05} .}
\label{fig:HDFS-galcounts}
\end{center}
\end{figure}

\begin{figure}
\begin{center}
\resizebox{\columnwidth}{!}{\includegraphics{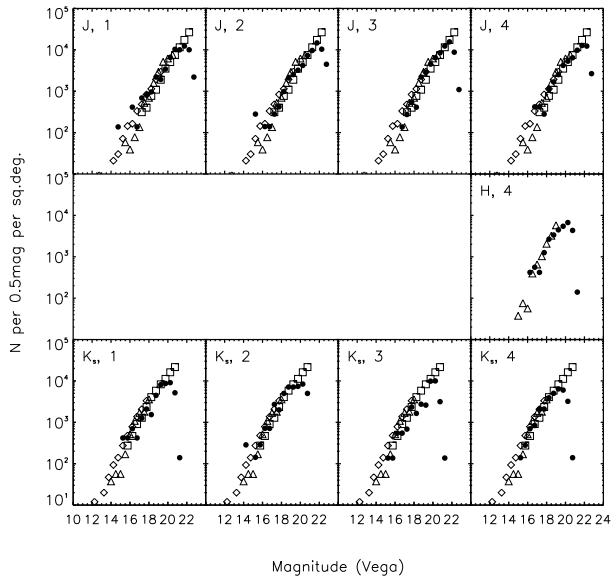}}
\caption{Same as Fig.~\ref{fig:HDFS-galcounts} but for the CDF-S region.}
\label{fig:CDFS-galcounts}
\end{center}
\end{figure}

Figs.~\ref{fig:HDFS-galcounts} and ~\ref{fig:CDFS-galcounts} show
the galaxy counts computed from the catalogs extracted from the
EIS-DEEP images of each field (columns) and passband (rows) of the
HDF-S and CDF-S regions, respectively. This is done in order to
minimize differences that could be introduced in mean counts from
field-to-field variations in depth of the observations. Here all
objects with stellarity index $< $ 0.95 were classified as
galaxies. The present counts are compared to those published
earlier by different authors \citep{vaisanen00,
martini01,iovino05}. The good agreement between the various counts
is further evidence of the scientific quality of the data and
calibration presented in this paper. One can also see the strong
field-to-field variations in limiting magnitude, a problem one
must take into account when dealing with mosaics covering large
areas of the sky.

\begin{figure}
\begin{center}
\resizebox{\columnwidth}{!}{\includegraphics{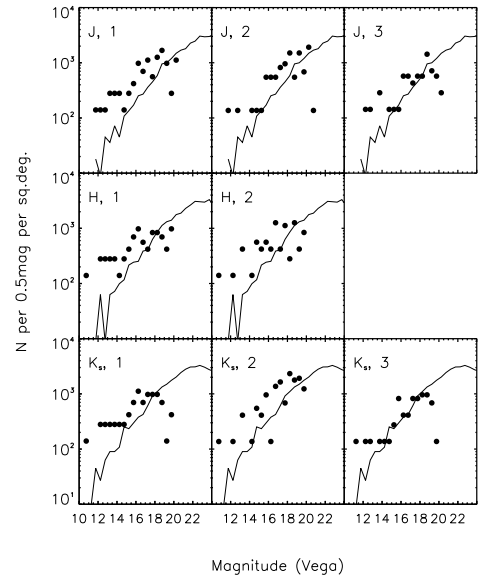}}
\caption{Star counts in the three passbands (rows) for the three fields (columns) covering HDF-S
(full circles) compared to the predictions (solid line) computed from
the model of \cite{girardi05}.}
\label{fig:HDFS-starcounts}
\end{center}
\end{figure}

\begin{figure}
\begin{center}
\resizebox{\columnwidth}{!}{\includegraphics{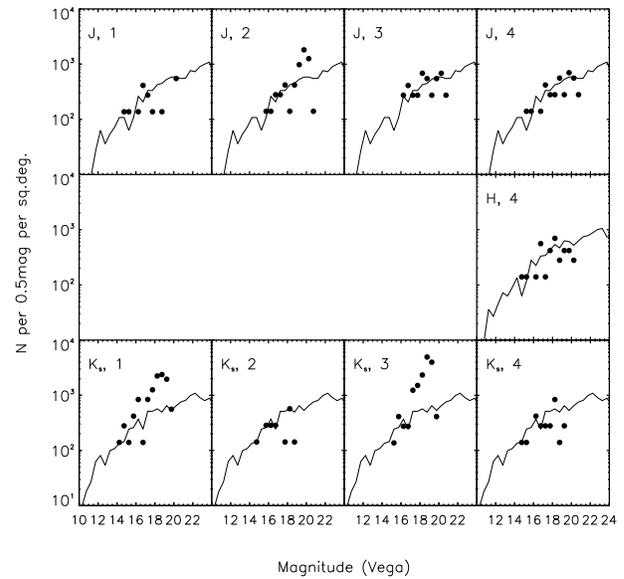}}
\caption{Same as Fig.~\ref{fig:HDFS-starcounts} but  for the four fields of  CDF-S}
\label{fig:CDFS-starcounts}
\end{center}
\end{figure}

For completeness, Figs. ~\ref{fig:HDFS-starcounts} and
~\ref{fig:CDFS-starcounts} show similar plots for the stellar
counts, comparing them with the predictions of the Galactic model
of \cite{girardi05}. Considering the uncertainties involved in the
model and the shot-noise the agreement is quite remarkable, with
the possible exception of the \k\ counts in the CDF-S-1 and -3.

\section{Summary}
\label{summary}

This paper presents the results of the infrared part of the EIS-DEEP
survey carried out in the period 1998-1999 covering the HDF-S and
CDF-S regions.  The paper includes re-processed and re-calibrated
images of previously released data, as well as new sets of data not
released before.  These data are available at the CDS\footnote{The
science grade images and catalogs are available at the CDS via
anonymous ftp to cdsarc.u-strasbg.fr (130.79.128.5) or via
http://cdsweb.u-strasbg.fr/cgi-bin/qcat?J/A+A/} and supersede earlier
releases described in the preprints by \cite{dacosta98, rengelink98}
and \cite{vandame01} which represent different phases of the project.
In contrast to the original release, new datasets have become
available, allowing a critical examination of the results by carrying
out an array of direct comparisons with reductions of the same dataset
as well as comparison with independent datasets.

The results show that the past problems with flux loss of {\it Jitter}
and an earlier version of the EIS/MVM library have been successfully
corrected with the magnitudes measured from the EIS images showing no
systematic deviations relative to those derived from images reduced
using other reduction techniques.  Furthermore, comparisons with data
obtained by other authors also suggest that the absolute calibration
is reliable to within $\sim5-10\%$. To obtain a better accuracy would
require additional observations under photometric conditions and a
good sampling of standards over the entire night.

The results obtained in this paper, combined with those discussed in
previous papers of this series, demonstrate the power and versatility
of the EIS Data Reduction System which can be used to efficiently
process optical and infrared data in the same environment yielding
reliable results. The EIS-DEEP survey was the precursor of the the
Deep Public Survey (DPS), which combines optical wide-field imaging
with infrared data. The results of this survey will be reported in
forthcoming papers of this series
\citep{olsen06b,mignano06}. Combined, these data provide information
in up to seven bands from the ultraviolet to the near-infrared, thus
being an excellent test case for future optical/infrared surveys.  The
results presented here and in other papers of this series demonstrate
that the EIS Data Reduction System meets the requirements for
long-term programs using the VST and VISTA survey telescopes.

\begin{acknowledgements}

We thank the anomymous referee for useful comments which improved the
manuscript.  This publication makes use of data products from the Two
Micron All Sky Survey, which is a joint project of the University of
Massachusetts and the Infrared Processing and Analysis
Center/California Institute of Technology, funded by the National
Aeronautics and Space Administration and the National Science
Foundation. This publication make use of the Guide Star Catalog, which
was produced at the Space Telescope Science Institute under
U.S. Government grant. These data are based on photographic data
obtained using the Oschin Schmidt Telescope on Palomar Mountain and
the UK Schmidt Telescope. We thank all of those directly or indirectly
involved in the EIS effort.  Our special thanks to M. Scodeggio for
his continuing assistance, A. Bijaoui for allowing us to use tools
developed by him and collaborators over the years and past EIS team
members for building the foundations of this program.  LFO
acknowledges financial support from the Carlsberg Foundation, the
Danish Natural Science Research Council and the Poincar\'e Fellowship
program at Observatoire de la C\^ote d'Azur.  The Dark Cosmology
Centre is funded by the Danish National Research Foundation. JMM would
like to thank the Observatoire de la C\^{o}te d'Azur for its hospitality
during the writing of this paper.

\end{acknowledgements}

\bibliographystyle{../../../aa}
\bibliography{/home/lisbeth/tex/lisbeth_ref}

\end{document}